# Stacking enabled strong coupling of atomic motion to interlayer excitons in van der Waals heterojunction photodiodes

**Authors:** Fatemeh Barati[1,2,†], Trevor B. Arp[1,2,†], Shanshan Su[3], Roger K. Lake[3], Vivek Aji[2], Rienk van Grondelle[4,5,*], Mark S. Rudner[6,*], Justin C.W. Song[7,*], Nathaniel M. Gabor[1,2,5*]

**Affiliations:**

[1]Laboratory of Quantum Materials Optoelectronics, University of California, Riverside, CA 92521, USA.

[2]Department of Physics and Astronomy, University of California, Riverside, CA 92521, USA.

[3]Laboratory for Terahertz and Terascale Electronics (LATTE), Department of Electrical and Computer Engineering, University of California - Riverside, Riverside, California 92521, USA.

[4]Department of Physics and Astronomy, Faculty of Sciences, Vrije Universiteit Amsterdam, De Boelelaan 1081, 1081 HV, Amsterdam, Nederland.

[5]Canadian Institute for Advanced Research, MaRS Centre West Tower, 661 University Avenue, Toronto, Ontario ON M5G 1M1, Canada.

[6] Niels Bohr Institute, University of Copenhagen, 2200 Copenhagen, Denmark.

[7]Division of Physics and Applied Physics, School of Physical and Mathematical Sciences, Nanyang Technological University, Singapore 637371.

*r.van.grondelle@vu.nl, rudner@nbi.ku.dk, justinsong@ntu.edu.sg, nathaniel.gabor@ucr.edu

† These authors contributed equally to this work

**Abstract:** We reveal stacking-induced strong coupling between atomic motion and interlayer excitons through photocurrent measurements of $WSe_2/MoSe_2$ heterojunction photodiodes. Strong coupling manifests as pronounced periodic sidebands in the photocurrent spectrum in frequency windows close to the interlayer exciton resonances. The sidebands, which repeat over large swathes of the interlayer exciton photocurrent spectrum, occur in energy increments corresponding directly to a prominent vibrational mode of the heterojunction. Such periodic patterns, together with interlayer photoconductance oscillations, vividly demonstrate the emergence of extraordinarily strong exciton-phonon coupling - and its impact on interlayer excitations - in stack-engineered van der Waals heterostructure devices. Our results establish photocurrent spectroscopy as a powerful tool for interrogating vibrational coupling to interlayer excitons and suggest an emerging strategy to control vibronic physics in the solid-state.

**One Sentence Summary:** Stacking van der Waals crystals significantly strengthens the interactions between atomic vibrational motion and interlayer excitons.

**Main Text:** Atoms - when combined into molecules and crystals - form a dynamic scaffolding that provides the energy landscape for electronic excited states. While atomic motion is ordinarily expected to dissipate energy from electronic excitations, strong coupling of vibrational motion to excited states can significantly transform light-matter interactions. In soft matter, such as photosynthetic light harvesting complexes, the interplay between atomic motion and exciton dynamics enhances electronic energy transfer[1-4] in spite of the fluctuating physical environment. In crystals, the presence of strong interactions between electronic excitations and phonons - the elementary excitations of the atomic lattice - warps optical line shapes[5,6], enables the trapping of excitations[7-9], allows mechanical control of electron transport[10-11], and drives the formation of exotic exciton-phonon quasiparticles and excitonic complexes[12,13].

Engineering the interactions between atomic motion in a crystal lattice and excitons is challenging since the interaction strength is largely fixed by its atomic-scale configuration. As a result, such exciton-phonon coupling is typically thought of as an *intrinsic* property. This coupling is most pronounced for localized excitations such as those in molecules and molecular crystals[7,9], where it gives rise to periodic structures in absorption spectra (Franck-Condon progressions). In contrast, strong coupling to atomic motion rapidly diminishes when excitons are delocalized over many atomic sites[9] and can move freely, as is the case in particular in monolayer van der Waals (vdW) crystals. Indeed, delocalized excitons found in individual semiconducting vdW monolayers display comparatively weak exciton-phonon coupling as indicated by broad and generally isolated absorption peaks[14-16].

Here we report the emergence of extraordinarily strong exciton-phonon coupling when individual vdW layers are combined to form atomically thin heterostructure devices. While excitons within the constituent layers (intralayer excitons) remain relatively unchanged, the interlayer excitons - with constituent electrons and holes in adjacent layers - become strongly affected by atomic vibrational motion. The interlayer photocurrent that results from liberating electron-hole (*e-h*) pairs from the bound interlayer exciton state exhibits rich dependence on photon energy $E_{PH}$, gate voltage $V_G$, and source drain voltage $V_{SD}$. Strong coupling manifests itself as numerous photocurrent sidebands that form a palisade of multiple vibrational excited state resonances above the lowest excited state absorption feature. Surprisingly, the effects of extreme exciton-phonon coupling go far beyond excited state absorption characteristics (*e-h* pair generation), giving rise to periodic oscillations within the optoelectronic transport characteristics (*e-h* pair separation) even at room temperature. These twin manifestations in both *e-h* pair generation and separation provides distinct snapshots of the unusually strong interactions that emerge between atomic motion and excitons in vdW heterostructures[17-20] and heralds a new regime of molecular-like vibronic excited states in solid state semiconductors.

We studied vdW heterostructures composed of bilayer tungsten diselenide ($WSe_2$) stacked on top of monolayer molybdenum diselenide ($MoSe_2$), as shown in Fig. 1. Using an inverted fabrication process (Supplementary Materials Section S1), we patterned multilayer graphene gate electrodes and conventional metal source and drain contacts. The $WSe_2$-$MoSe_2$ heterostructures - assembled and characterized independently - were laminated onto the prefabricated device patterns. Combining the vdW heterostructure in this way enabled complete protective encapsulation of the $WSe_2$-$MoSe_2$ interface region between hexagonal boron nitride layers (Fig. 1A, 1B).

Devices were characterized using Raman, photoluminescence (PL), and photocurrent (PC) spectroscopy. As shown in Fig. 1C, the Raman spectrum of the heterostructure (black line) exhibits peaks that are also evident in the $MoSe_2$ (green line) and $WSe_2$ (blue line). When comparing

Raman peaks in the heterostructure region to those in the constituent layers, we observed negligible shift of the peak positions as a function of energy. Importantly, however, we observed the highest intensity peaks at the WSe$_2$-MoSe$_2$ interface (black line), where the most prevalent features in the Raman spectrum occur at energies near 30 meV (29.9 meV, 31.0 meV, and 31.9 meV, Fig. 1C inset), mirroring those of the constituent layers.

Fig. 1D compares the PL vs. photon energy $E_{PH}$ from MoSe$_2$ (green line), WSe$_2$ (blue line), and the heterostructure region (black line). The heterostructure exhibits similar spectral features to the individual layers, consistent with ordinary behavior when stacking vdW heterostructures[21-26]. Due to charge redistribution at the interface, low-energy bound $e$-$h$ pairs (excitons) form between the valence band of bilayer WSe$_2$ and the conduction band of MoSe$_2$ (Fig. 1B). Importantly, direct PL signatures of the interlayer excitons are washed out by their very small oscillator strength; nevertheless, detailed temperature dependent measurements of the excess current due to hot carrier excitation indicated that the energy required to excite interlayer $e$-$h$ pairs is $E_I \sim 1.0$ eV[27], corresponding to near infrared wavelengths.

In order to overcome the small oscillator strength that prevents strong photoluminescence from interlayer excitons, we instead employed measurements of the interlayer photocurrent $I_{PC}$ with the laser focused on the WSe$_2$-MoSe$_2$ interface. Sweeping $E_{PH}$, we find that $I_{PC}$ is particularly pronounced in two frequency windows occurring near $E_{PH} \sim 1.3$ eV and $\sim 0.9$ eV (for details see Supplementary Materials Section S2.1 and S4.2). Fig. 2 examines the detailed dependence of this photocurrent on $E_{PH}$ and $V_G$. Excitation energies in the vicinity of these two hot spots correspond to interlayer excitons hosted in the WSe$_2$-MoSe$_2$ heterostructure, namely the $K \rightarrow K$ ($\sim 1.3$ eV) and $\Gamma \rightarrow K$ ($\sim 0.9$ eV) interlayer excitons[21,27-28].

Intriguingly, when $E_{PH}$ was tuned between 1.24 eV and 1.40 eV (Fig. 2A), we observed a periodic sequence of photocurrent peaks that occur in a narrow range of gate voltages (near $V_G = -3.5$ V). While the strongest peak occurs at $E_{PH} = 1.32$ eV, it is only slightly stronger than several equally spaced maxima at higher and lower $E_{PH}$ (Fig. 2B). To confirm this periodic modulation, we calculated the Fourier transform of the second derivative of the photocurrent data, where we find a clear periodic component at $1/\Delta\varepsilon = (30$ meV$)^{-1}$ (blue dashed line Fig. 2c). Interestingly, near the interlayer excitation from the WSe$_2$ $K$-valley to the MoSe$_2$ $K$-valley (Fig. 2D), the discrete energy difference $\Delta\varepsilon = 30$ meV between photocurrent peaks closely corresponds to the strongest phonon mode in the heterojunction, $\hbar\Omega \sim 30$ meV, as characterized by the Raman spectrum (see Fig. 1C). Here $\hbar$ is Planck's constant and $\Omega$ is the phonon frequency.

Strikingly, optical excitation of the lowest lying ($\Gamma \rightarrow K$) interlayer exciton also results in a series of approximately equally spaced discrete sidebands that match the energy of the prevailing vibrational modes. This behavior is highlighted in Fig. 2E, which shows $I_{PC}$ vs. $E_{PH}$ and $V_G$ at infrared photon energies. In the range $E_{PH} = 0.88 - 1.03$ eV, we observed a set of evenly spaced photocurrent maxima, which increase in amplitude as $E_{PH}$ increases. The lowest energy peak occurs at $E_{PH} = 0.90$ eV, and line traces of $I_{PC}$ vs. $E_{PH}$ (Fig. 2F) show regularly spaced peaks that are superimposed on a photocurrent background that increases with $E_{PH}$. Taking the Fourier transform of the second derivative of the interlayer photocurrent data (Fig. 2G) reveals two periodic components: a dominant component at $\Delta\varepsilon = 30$ meV and a weaker component at 22 meV.

The appearance of a periodic array of photocurrent sidebands can be understood through strong coupling between vibrations of the atomic lattice and interlayer excitons (Fig. 3). When excitons are delocalized - moving freely through the lattice - and only weakly coupled to the displacements of the atoms, the total energy of the system is approximately given by a sum of independent electronic and mechanical (lattice) contributions. In this case, as shown in Fig. 3A, the energies of the electronic ground and excited state manifolds shift equally and remain parallel as a function of the atomic displacement coordinate (shown schematically as the horizontal axis in Fig. 3). Here, the lowest-energy electronic excited state corresponds to a delocalized exciton on top of an un-distorted lattice. In this situation, a single, isolated absorption peak is expected (wide blue band in Fig. 3A, right).

When the crystal lattice and excitons are strongly coupled, the interatomic potential is distorted by the electronic excited state. In this case, the lowest energy electronic excited state may be that of a *localized* exciton, accompanied by a local displacement of the lattice (Fig. 3B): within the electronic excited state manifold, the minimum of the effective interatomic potential is displaced relative to its location for the electronic ground state. Due to this displacement, finite overlaps develop between the ground state phonon wavefunction and multiple phononic excited states in the electronic excited state manifold. Due to the fact that the electronic excitations occur far faster than lattice relaxational dynamics, these overlaps give rise to multiple Franck-Condon sidebands in the absorption spectrum (right panel) (see extended discussion in the Supplementary Materials Section S5). When the (sideband) generated *e-h* pairs are separated, they form the numerous and periodic photocurrent sidebands in increments of the phonon energy $\hbar\Omega$ observed in Fig. 2. These sidebands vividly demonstrate electronic transitions that depend intimately on the phonon coordinates.

To explore the consequences of such extreme coupling in the separation of *e-h* pairs, we examined the heterojunction current-voltage characteristics. Discretely spaced peaks in the photocurrent spectra (Fig. 2) emerge only if the device is tuned to charge neutrality, the condition at which electrons and holes across the interface are fully compensated (Fig. 1B, Supplementary Materials Section S4.3). At charge neutrality, the $I$-$V_{SD}$ characteristics are highly asymmetric (Fig. 4A): the dark current increases exponentially in forward bias ($V_{SD} > 0$ V) and is suppressed in reverse bias ($V_{SD} < 0$ V), consistent with ordinary vdW heterostructure p-n junction behavior [17,29,30]. To compare to ordinary heterojunction devices, we fit the dark $I$-$V_{SD}$ characteristic to the usual diode form $I = I_0(\exp(eV_{SD}/\alpha k_B T) - 1))$, where $I_0$ is the reverse bias saturation current, $k_B T = 26$ meV is the thermal energy at room temperature, and the factor $\alpha \geq 1$ is a phenomenological constant that relates changes in the interlayer electric field energy to a reduced voltage $eV_{SD}/\alpha$ [27,31]. The diode equation (solid line in Fig. 4A) shows excellent agreement with the data, yielding an $\alpha = 1.82$ fit.

Remarkably, the reverse bias photoconductance $dI_{PC}/dV_{SD}$ exhibits clear voltage-dependent oscillations in increments of $\hbar\Omega \sim 30$ meV, underscoring the importance of strong coupling in the *e-h* pair separation process (Fig. 4A inset). The change in electric field energy due to an effective interlayer voltage $V_I$ can be written as $\Delta E = \vec{p} \cdot \vec{E} = ed\, V_I/\eta t_{TMD}$, where $|\vec{p}| = ed$ is the dipole moment of the interlayer exciton with length $d$, $t_{TMD}$ is the heterostructure thickness, and $\eta > 1$ reduces the effective voltage due to diode junction capacitance. Using the dependence of the photocurrent on $V_G$ and $V_{SD}$ to extract $\eta = 2.23$ (Supplementary Materials Section S4.4) and approximating $d \sim t_{TMD}$, we are able to relate changes in the interlayer electric field energy to a reduced voltage $eV_{SD}/\alpha$, giving $\Delta E \approx e V_I/\eta = e V_{SD}/\alpha\eta$. Using the relationship $\Delta E = e V_{SD}/\alpha\eta$ to rescale $V_{SD}$, we can now plot the derivative of the photoconductance $d^2 I_{PC}/d\Delta E^2$ (Fig. 4B) to obtain the period of oscillations. From the Fourier transform of the data (Fig. 4C), we observe voltage-dependent oscillations with a period that precisely matches the strongest phonon mode.

Strong exciton-phonon coupling and multiple sidebands in the *e-h* pair generation process are hallmarks of localized excitons. In contrast to delocalized excitons (Fig. 3A), localized excitons (gray curve Fig. 3B) readily take on a lower energy state by re-arranging the local lattice structure (red curve Fig. 3B)[7,9], resulting in a displaced potential well (red curve) that produces multiple sidebands in the excitation process (Supplementary Materials Section S5). The numerous phonon sidebands observed in Fig. 2 indicate that the well is deep; taking the midpoint of the sideband oscillations as a rough gauge, we estimate that the well lowers the excited state energy by $\sim 3 \times \hbar\Omega \approx 90$ meV. Because the depth of the well is significantly larger than the available thermal energy $k_B T = 26$ meV at $T \sim 300$ K, the sidebands persist even at room temperature (Fig. 4). While there are many possible mechanisms for exciton localization, among the most tantalizing is the self-trapping of excitons through strong interactions with the phonon field[7,9,12]. Indeed, the sheer depth of the well $\sim 90$ meV, which is comparable to the interlayer exciton binding energy $\sim 150$ meV[32], could readily provide a self-trapping potential that localizes the exciton.

Strikingly, even as the interlayer exciton exhibits signatures of localization, the strong interlayer photocurrent confirms that, once separated, electrons and holes that make up the interlayer exciton are delocalized and move freely in the conduction (valence) bands of $WSe_2/MoSe_2$. Further, while the intralayer excitons (at excitation energies near the intralayer PL emission) also produce photocurrent, they do not exhibit the sidebands of Fig. 2 nor the oscillations of Fig. 4 (Supplementary Materials Section S4.1). These show that signatures of localization manifest for the interlayer excitons amongst a background of delocalized excitations. This singles out interlayer excitons as particularly sensitive to exciton-phonon coupling that emerges from the stacking of two vdW crystal layers.

Stack-engineering - exemplified here by stacking two vdW layers $MoSe_2$ and $WSe_2$ - can be used to unlock strong exciton-phonon coupling and, concomitantly, to engineer opto-mechanical interactions. This closely parallels recent efforts in vdW layer engineering that have enabled electrical control of their stacking configuration[33,34]. From a broader perspective, the strong exciton-phonon coupling reported here mirrors that of vibronic states that enable phenomena ranging from singlet fission[35] and exciton dissociation[36] in organic compounds to long-lived coherent dynamics and enhancements to exciton transport in photosynthetic complexes[1-4,37]. Our findings will inspire comprehensive measurements aimed at establishing whether quantum coherent mixing of exciton and charge transfer states - as recently observed in the photosystem II reaction center - could be exploited for ultra-efficient solar energy harvesting[37-39] in a solid-state setting. We anticipate that stack-engineering of strong exciton-phonon coupling will establish vdW heterojunctions as a versatile platform for controlling vibronic-like physics in 2D semiconductors devices.

**Acknowledgements:** The authors would like to acknowledge valuable discussions with Vasili Perebeinos. This work was supported by the Presidential Early Career Award for Scientists and Engineers (PECASE) through the Air Force Office of Scientific Research award no. FA9550-20-1-0097 (N.M.G, T.B.A.), through support from the National Science Foundation Division of Materials Research CAREER award no. 1651247 (N.M.G., F.B.), and through the United States Department of the Navy Historically Black Colleges, Universities and Minority Serving Institutions (HBCU/MI) award no. N00014-19-1-2574 (N.M.G., F.B.). N.M.G. acknowledges support through the Canadian Institute for Advanced Research (CIFAR) Azrieli Global Scholar Award. T.B.A. acknowledges support from the Fellowships and Internships in Extremely Large Data Sets (FIELDS) program, a NASA MUREP Institutional Research Opportunity (MIRO) program, grant number NNX15AP99A. J.C.W.S. acknowledges support from the National Research Foundation (NRF), Singapore under its NRF fellowship programme award number NRF-NRFF2016-05, the Singapore Ministry of Education under its MOE AcRF Tier 3 Award MOE2018-T3- 1-002. M.S.R is grateful for the support of the European Research Council (ERC) under the European Union Horizon 2020 Research and Innovation Programme (Grant Agreement No. 678862), and the Villum Foundation. R.K.L and S.S. acknowledge support from the NSF EFRI-1433395. This work used the Extreme Science and Engineering Discovery Environment (XSEDE), which is supported by National Science Foundation grant number ACI-1053575 and allocation ID TG-DMR130081.

**Author Contributions:** F.B. and T.B.A. contributed equally to this work, performing device fabrication, detailed experiments, analysis and modelling. N.M.G. conceived the experiments, as well as supervised the experimental, theoretical, and computational components with additional input from R.L, V.A., R.vG, M.S.R. and J.C.W.S. S.S. and R.L. conducted detailed computations of the vibrational band structure to support the interpretation. V.A., M.S.R. and J.C.W.S. provided theoretical support, while R.vG. advised experimental concepts. All authors contributed to the writing of the manuscript.

**Competing Interests:** Authors declare no competing interests.

**Data and materials availability:** The code that generates Fig. 1, 2, and 4 and all supplementary figures that present additional data is published alongside this work. The code can be found at https://github.com/qmolabucr/wse2mose2 and this repository includes all relevant data such that the results can be fully replicated[40].

**Supplementary Materials:**

Materials and Methods

Supplementary Text

Figures S1.1-S4.10

References 41-58

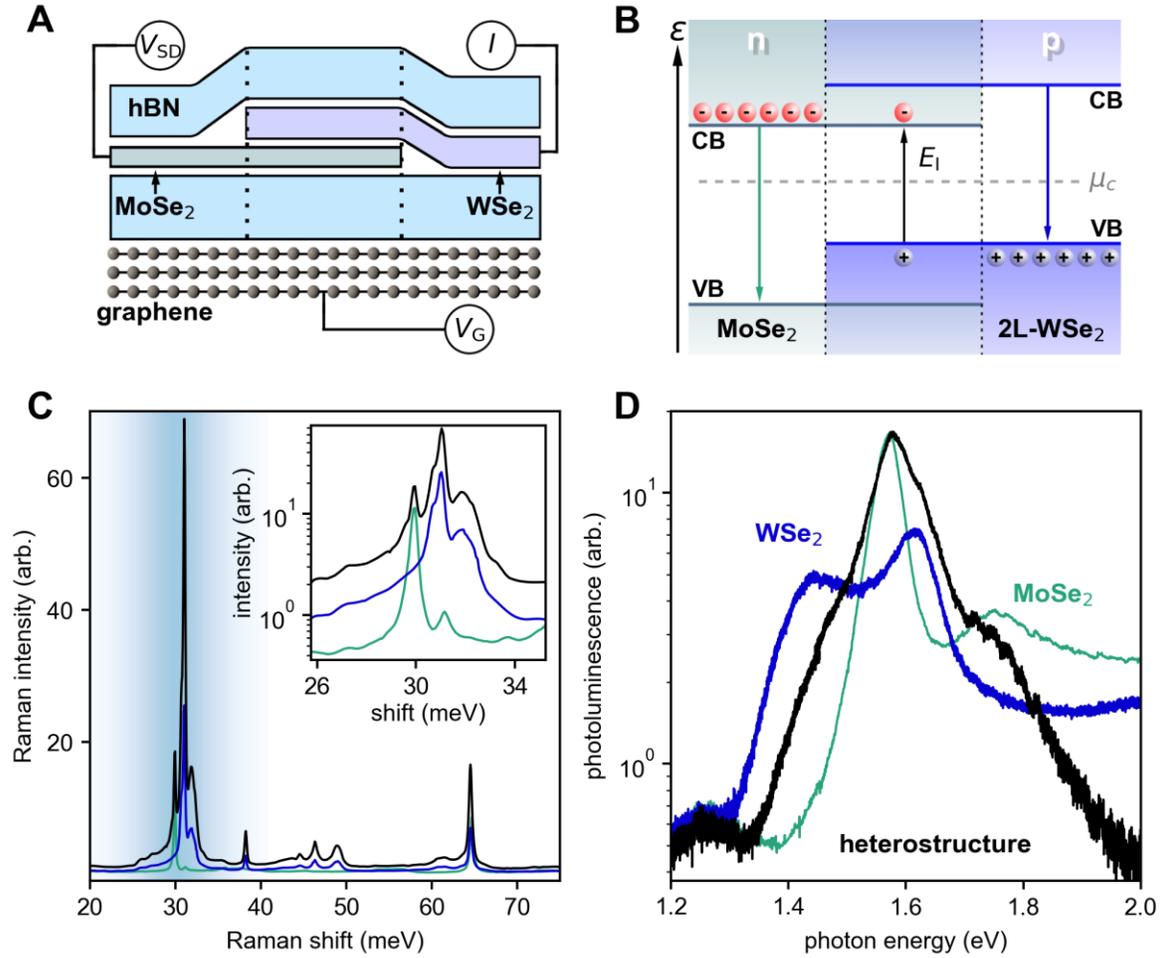

**Fig. 1.** Optical characterization of encapsulated van der Waals p-n heterojunction devices. (**A**) Schematic of the hexagonal boron nitride (hBN) encapsulated heterostructure with a multilayer graphene back gate and source-drain electrodes (heterostructure overlap area ~10 $\mu m^2$). (**B**) Electronic energy band diagram at $V_{SD} = 0$ V, showing the conduction bands (CB), valence bands (VB), and chemical potential $\mu_c$ in equilibrium (horizontal dashed gray line). (**C**) Raman spectroscopy ($\lambda = 532$ nm) of MoSe$_2$ (green), bilayer WSe$_2$ (blue), and the heterostructure (black). **inset**, log scale Raman intensity versus Raman shift near 30 meV. (**D**) Photoluminescence ($\lambda = 532$ nm) from MoSe$_2$ (green), bilayer WSe$_2$ (blue), and the heterostructure (black).

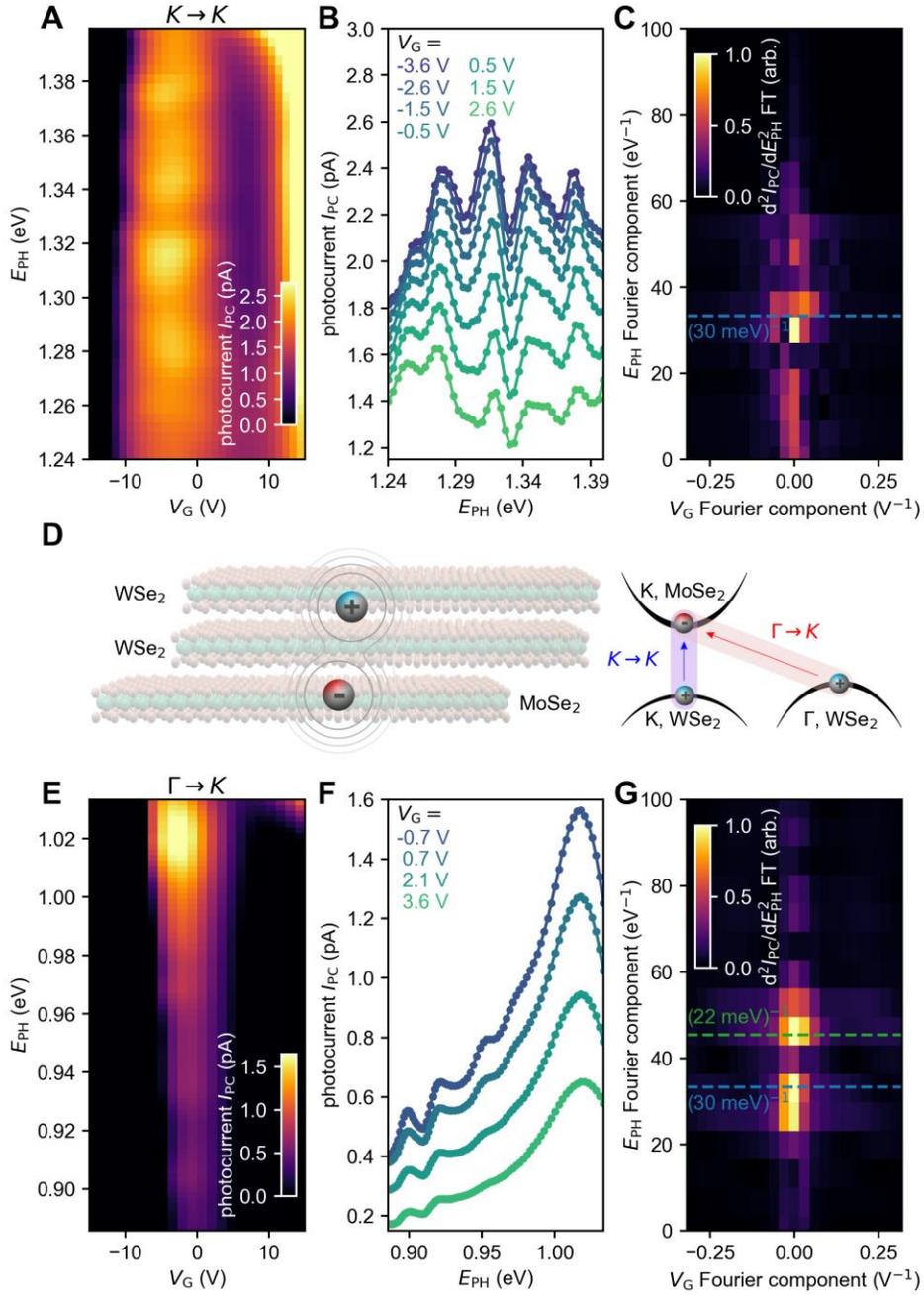

**Fig. 2.** Multiple and periodic sidebands in the photocurrent spectra of van der Waals p-n heterojunction devices. (**A**) Interlayer photocurrent vs. $E_{PH}$ and $V_G$ of the heterostructure near the direct ($K \rightarrow K$) interlayer exciton transition at $T$ = 20 K. (**B**) Line cuts of the photocurrent spectrum. Some lines removed for clarity, for full data see Supplementary Materials Section S4.2. (**C**) 2D-Fourier transform of the photocurrent second derivative. (**D**) Left, schematic of the interlayer exciton in the layered structure, and right, a schematic of the band structure with the momentum direct ($K \rightarrow K$) and indirect ($\Gamma \rightarrow K$) interlayer excitons labeled. (**E**) $I_{PC}$ vs. $E_{PH}$ and $V_G$ of the heterostructure near the $\Gamma \rightarrow K$ exciton at $T$ = 20 K. (**F**) Line cuts of the photocurrent spectrum and (**G**) 2D-Fourier transform of the photocurrent second derivative.

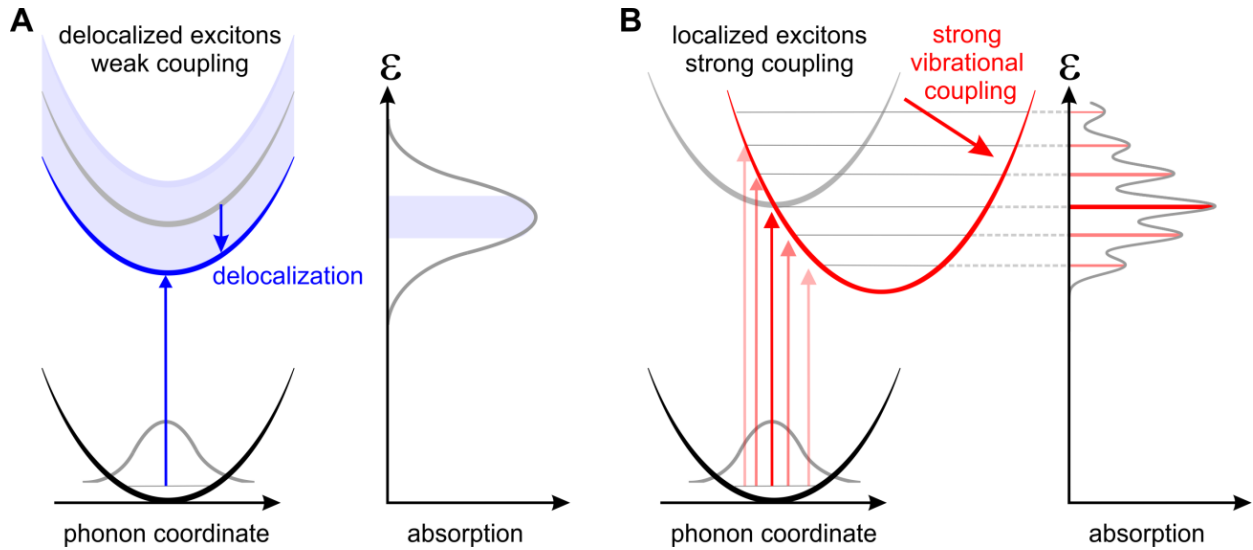

**Fig. 3.** Localization of interlayer excitons due to strong exciton-phonon coupling in van der Waals heterostructures. (**A**) Delocalized exciton band structure in the presence of weak coupling to atomic motion. Delocalization forms a band of states (blue), allowing minimization of the total energy of the exciton, and leads to a broad absorption peak. (**B**) Localized exciton band structure in the presence of strong coupling to atomic motion. Grey parabola represents the bare exciton band. Strong coupling results in a large shift in vibrational coordinates in the excited state (red). Absorption then exhibits a set of discrete sidebands. In $WSe_2$-$MoSe_2$ heterostructures, interlayer excitons are comprised of electrons and holes in different layers and misaligned valleys. As a result, direct excitation of highly delocalized modes such as a zero center of mass momentum mode (thick blue line, panel A) is suppressed. Instead, excitation of localized interlayer excitons, which are strongly coupled to phonons, dominates (panel B) as demonstrated by photocurrent sidebands in Fig. 2.

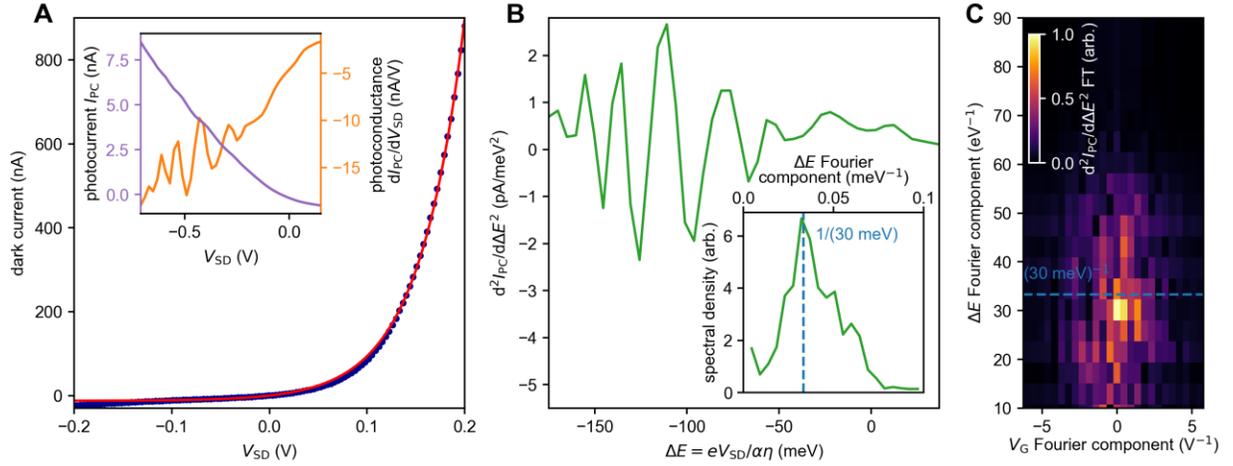

**Fig. 4.** Optoelectronic transport characteristics of the van der Waals p-n heterojunction at room temperature. (**A**) Interlayer dark current vs. $V_{SD}$ (blue points) and fit to the diode equation (red curve). **A, inset**, Photocurrent, photoconductance extracted from a large set of spatial images taken near the charge neutrality region between $V_G$ = -0.3 and 0.4 V, $E_{PH}$ = 0.99 eV. (**B**) Second derivative of heterostructure photocurrent vs. $V_{SD}$ scaled by factors $\alpha$ and $\eta$ to give the total interlayer energy difference $\Delta E$. (**B**) **inset**, Fourier transform of the data in B. (**C**) 2D-Fourier transform of the second derivative of the heterostructure photocurrent versus $V_G$. For full data and analysis see Supplementary Materials Sections S2.2 and S4.3.

**Supplementary Materials:**

**Section S1. WSe$_2$/MoSe$_2$ Device Fabrication**

In this work, we present intensive spectroscopic, electronic, and optoelectronic characterization of precisely fabricated heterostructure devices composed of bilayer WSe$_2$ and monolayer MoSe$_2$. Based on our previous temperature dependent electronic transport study on the same material system[27], this unique combination of materials was chosen due to its indirect band gap, unique phonon band structure, and long-lived interlayer excitons expected at infrared wavelengths.

The heterostructure devices are assembled using a highly customized, temperature-controlled transfer microscope that ensures the interface between the two layers has no intentional contact to polymer films. The first step of the fabrication is to make the back-gate assembly as shown in Fig. S1B. The back gate consists of a layer of graphite (2 nm thick) as the conductor and a hexagonal boron nitride (h-BN) layer (12 nm) on top as the gate dielectric. The graphite layer is picked up using the h-BN and mounted on the prefabricated gate contacts. Fig. S1A shows the device geometry with all components in two different perspectives. Fig. S1B shows optical images of the three stages of the fabrication process from left to right. The left image in Fig. S1B shows the first step of the fabrication which is the gate assembly on the SiO$_2$ substrate. The next step is to write source-drain contacts on top of the bottom h-BN layer, these layers are labeled as MoSe$_2$ contact and WSe$_2$ contact shown in Fig. S1B middle. These contacts provide electrical access to each layer of the stack. For device 1 the MoSe$_2$/WSe$_2$ heterostructure was fabricated using mechanical exfoliation of WSe$_2$ and MoSe$_2$ flakes from bulk crystal (2D semiconductors) onto Si wafers coated with 290-nm-thick SiO$_2$. The silicon substrate is implemented as a back gate to tune the carrier density at the interface. Electrical access to the two layers of MoSe$_2$ and WSe$_2$ is achieved through independent electrical contacts to each material. The details of the fabrication process can be found in our previous study[27].

To minimize interfacial contamination, the two layers of the stack (WSe$_2$ and MoSe$_2$) are picked up via the top h-BN layer one after another in a two-step transferring process and mounted on the source drain contacts on the prefabricated gate. To get rid of possible impurities and destructive bubbles formed in between the layers of the heterostructure the device is finally annealed at 250 °C under Ar gas for one hour. The final step is to mount the device into a chip carrier within the measurement setup. The chip carrier is a patterned substrate of sapphire with gold for electrical access to the device. The electronic contacts on the SiO$_2$ substrate are wire bonded into the conductive pads of the chip carrier shown in Fig. S1 (left). The device is then soldered to the electronic pads of the cryostat and silver painted for further electronic experiments shown in Fig. S1C (right).

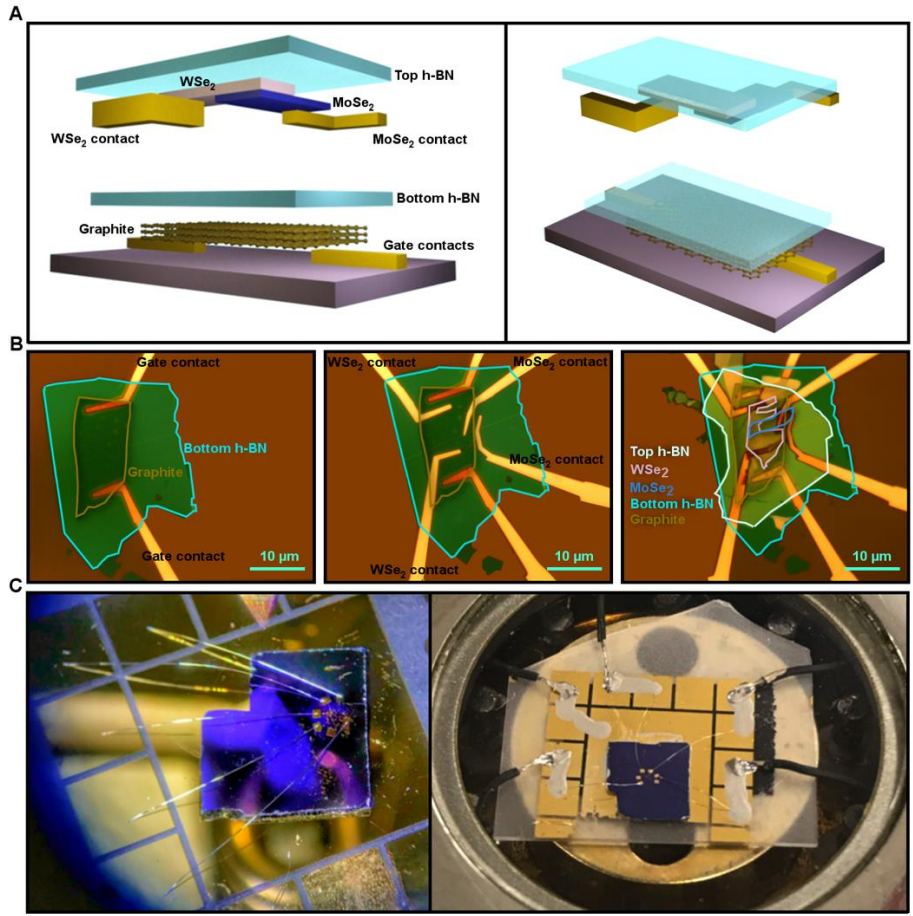

**Fig. S1.1** (**A**)**,** Two different perspectives of the device geometry, components labeled. (**B**)**,** Optical microscope images of the device showing the three-step fabrication process. Back-gate assembly consisting of electrical contacts made of gold, a layer of graphite and a layer of h-BN respectively from bottom to top. (left). Source-drain contacts are then drawn on top of the bottom h-BN layer (middle) for electrical access to the device. A stack of h-BN/WSe$_2$/MoSe$_2$ is then landed on these prefabricated electrical contacts (right). (**C**), Photographs of the devices wire-bonded into chip carriers. Device 2 (left) was contacted using a probe station cryostat (see Section S2.2). Device 1 (right) was contacted by soldering chip carrier to electrical pins of the cryostat (see section S2.1).

**Section S2.1. The Supercontinuum Scanning Photocurrent Spectroscopy Microscope**

The supercontinuum scanning photocurrent spectroscopy microscope (SSPSM) is a powerful tool designed to explore quantum optoelectronic properties of atomic scale devices by integrating optics and electronics into a combined quantum transport microscope. The SSPSM is equipped with a helium flow optical cryostat that enables both electrical access and optical illumination to nanoscale devices through electrical feedthroughs and a transparent window. This cryostat provides a stable and controllable low-pressure environment via a turbo pump system and can be cooled down to temperatures of $T = 4$ K.

The SSPSM microscope employs a supercontinuum Fianuim white laser as the light source along with Princeton Instrument monochromator for spatially and spectrally resolved supercontinuum excitations. The light source we use for this microscope is a Fianium WhiteLaseTM SC400-8; a picosecond pulsed laser pump (repetition rate = 20 MHz, pulse width = 6 ps) is coupled into a micro-structured optical fiber (the non-linear medium) of 1.5-meter length. At the output of the fiber, the collimated white beam of light has an average power of 8 W and a power spectral density of 4 mW/nm that ranges from $\lambda \sim 400\text{-}2200$ nm. Fig. S2.1 illustrates a schematic diagram of the microscope. The light that reaches the monochromator through the left optical line can be used for spatially and spectrally resolved laser excitations. The monochromatic laser light then generates a diffraction limited beam spot by passing through different optical components[41].

To probe the optoelectronic response of the samples we use scanning beam photoexcitation, as shown in Fig, S2.1 (GVS, L7 and L8) to generate a collimated beam of light on the back aperture of an objective. As the angle of the collimated beam is shifted the resulting diffraction limited beam spot that is scanned over the nanoscale device. The resulting current is amplified using a sub-100 fA resolution pre-amplifier over long averaging times (10-1000 ms) at low optical power. The reflected light from the sample transmits through the BS and reflects to a lens (L9) and gets collected by a photodiode detector (PD). The photodiode converts the optical power to electrical power as a photovoltage; the intensity of the photovoltage is monitored to form a simultaneous reflection image of the sample. To monitor and calibrate the output of the light source over the whole wavelength range, we use two photodetectors. A Si switchable gain detector (Thorlabs: PDA36A) is sensitive to the light from UV to NIR wavelength range (350-1100 nm), and an InGaAs photodetector (Thorlabs: DET10C) on the other hand is designed to detect light signals ranging from 900 to 1700 nm.

To achieve the highest performance of this microscope over the broad spectral range (400-2200 nm), optical components must match the wavelength range that we choose to conduct experiments. There are three different beam splitter coatings to cover the whole wavelength range. Thorlabs non- polarizing achromatic beam splitter cubes can provide a uniform reflection in these three ranges: BS028 for Vis (400-700 nm), BS029 for NIR (700-1100 nm), and BS030 for IR (1000-1600 nm). The beam splitters used for this microscope reflect 90% of the light to the sample; the 10% of the light that transmits through the BS can get detected with a power sensor (PS) coupled to a power meter (PM). The power sensor response function should also match the wavelength range of the microscope; using two power sensors (Thorlabs: S130C, and S132C) we can accurately detect the optical power in the ranges of 400-1100 nm and 700-1800 nm respectively in the power range of 500 pW-500 mW.

Data, in the form of two-dimensional maps of photocurrent and reflection, is acquired and displayed using python based custom control software[42]. Using the SSPSM microscope, we can perform spatial imaging as a function of wavelength by fixing the spectrum at a specific wavelength and conducting 2D spatial photocurrent mappings over the surface of the sample, while changing other parameters such as $V_{SD}$, $V_G$, or temperature in order to make 3D or 4D data cubes. We can also perform spectral imaging with a narrow resolution of 1 nm by fixing the laser beam at a specific spot on the device (e.g., the heterostructure as identified by correlating scanned reflection images to know device geometries) while tuning for example, the excitation wavelength. Additional parameters such as $V_{SD}$, $V_G$, or temperature are then tuned to form a multi-dimensional data set.

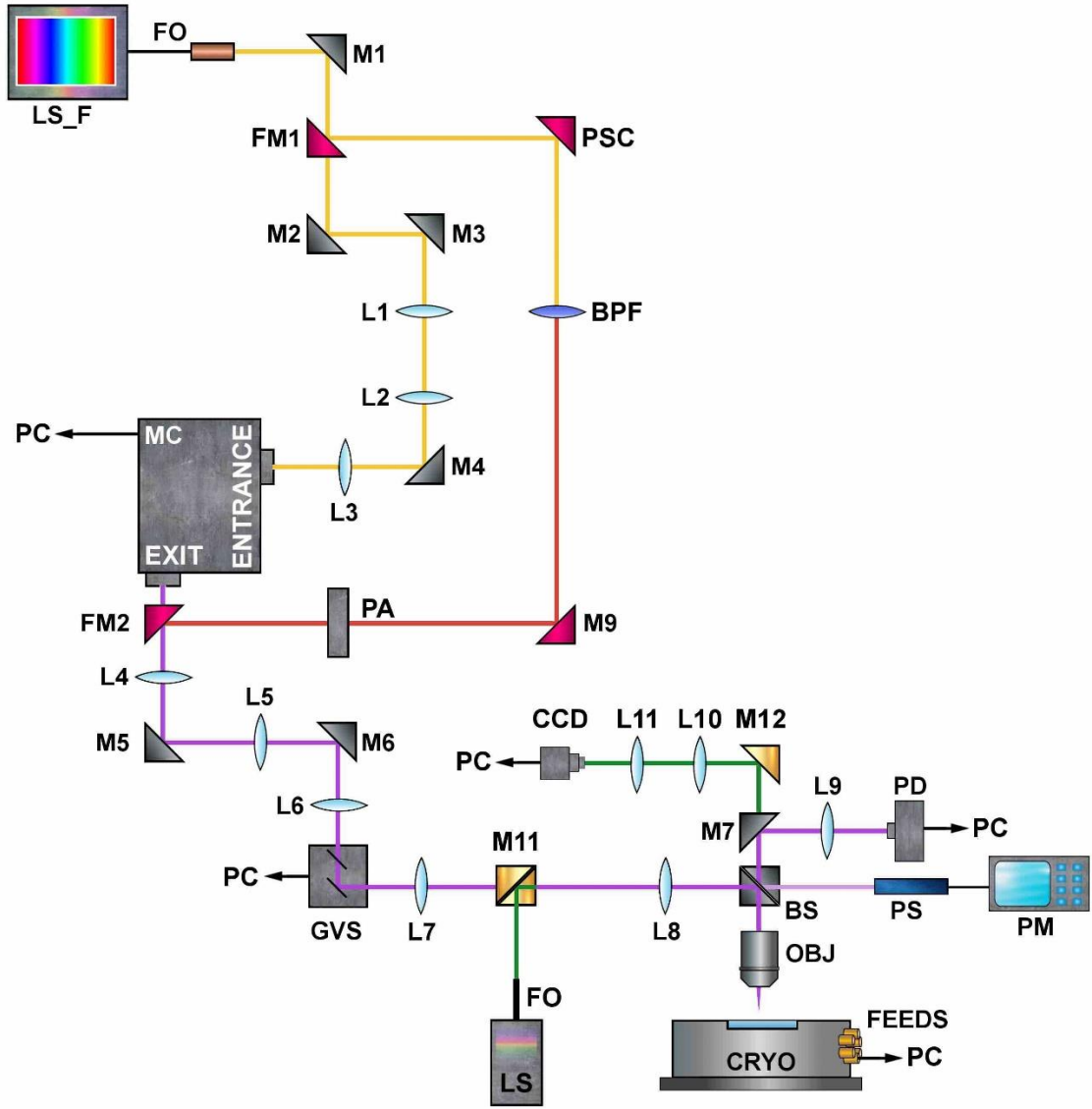

**Fig. S2.1** Schematic diagram of the supercontinuum scanning photocurrent spectroscopy microscope (SSPSM). Constituent components of the microscope comprise: light source of Fianium (LS_F), FO (fiber optics), monochromator (MC), mirrors (M), lenses (L), flip mount mirrors (FM), galvos system (GVS), beam splitter (BS), objective (OBJ), power sensor (PS), power meter (PM), photodiode detector (PD), CCD camera (CCD), cryostat (CRYO), PSC (periscope), band pass filter (BPF). Electronic experiments are conducted through the optical CRYO, while optical illumination is introduced via OBJ to devices (*41*).

**Section S2.2. Multi-Parameter Dynamic Photoresponse Microscopy Measurement Details**

The data intensive imaging of the sample phase space, shown in Fig. 4 was performed using the Multi-Parameter Dynamic Photoresponse Microscopy (MPDPM) technique described previously by Arp & Gabor[42]. Below we give a basic description of the most relevant aspects of the optics to this experiment, but more details on the hardware and methodology can be found in ref. 42. Data shown in the main figures is condensed from the MPDPM data analysis process, while the full imaging data sets can be found in section S4.3.

The core of the optical system is a Coherent MIRA 900 OPO ultrafast pulsed laser which generates 150 fs pulses with controllable wavelength from 1150 nm to 1550 nm at a 75 MHz repetition rate. The output of the laser is attenuated to control the laser power and fed into a scanning beam system similar to GVS, L7 and L8 in Fig. S2.1. After the second lens (L8 in Fig. S2.1) the MPDPM optics differ significantly from the SSPSM setup in order to get a stronger optical probe. In MPDPM a strong and highly local optical probe is required for best results, therefore the output of the scanning optics is directed into a GRIN (Gradient Index of Refraction) lens, which performs a similar function to the objective lens but with dramatically lower pulse dispersion. Combined with the fact that the OPO laser output has significantly higher power than the output of the SSPSM monochromator (order of mW at the sample) the pulse fluence is several orders of magnitude larger in the MPDPM setup than the SSPSM.

Photocurrent was measured using a lock-in amplifier. The laser beam was modulated by an optical chopper and the signal was passed through a pre-amplifier then locked-in to the chopper frequency. This was necessary to remove the substantial dark current that occurs due to the sample diode characteristics (see Fig. 4a). The data was then acquired through a National Instruments PCIe-6323 Data Acquisition Card. In addition to the photocurrent signal, the reflected light from the sample was continuously acquired along with the laser power.

The imaging data set shown in Fig S4.3, along with several similar image sets, was acquired using the "datacube" method from MPDPM, i.e. automatically taking a series of 2D scanning images as a function of one variable (with all other variables held constant) to form a three-dimensional data set, and then taking a series of these datacubes incremented over variable to get a 4-dimensional data set. Each set involved hundreds to thousands of images measured fully automatically over the course of several hours to a few days. For these longer measurements the main concern is laser stability, as any variation in the optics will cause the image location to drift.

To counter this, careful attention was paid to the alignment of the optics to avoid any geometrical aberration and the laser was run continuously for several days prior to any long imaging measurement to stabilize its performance. When the system is stable the main source of drift is thermal expansion of the various optical components; therefore, temperature and airflow of the lab is carefully maintained during measurements.

Even with the extra considerations for stability there will inevitably be some drift in the images, which must be corrected for in the data. There are multiple options but given that the laser power was monitored and shown to be constant throughout the measurement and the drift is small, the best method was to re-orient the data based on the reflection image. To find the correction for a given reflection image we use an algorithm that shifts the image by an integer number of pixels and calculates the difference between that image and the first image in the data set. By brute force the algorithm determines the shift that produces the minimum difference and uses that as the correction. This allows for correction that is accurate to the pixel resolution. For computational simplicity (and because the drift within datacubes was sub-pixel) we perform this procedure by comparing the first image in each 3D datacube to determine the shift for the whole cube.

Fig. S2.2 shows the drift correction over time for the longest measurement taken on this sample, which spanned 53 hours. In Fig. S2.2A we see the first and last images in the data set, and then the last image with the correction applied. Applying the correction removes some information near the edges, however the images were taken with a significant margin around the active area of the sample to prevent the drift correction from removing meaningful information. Fig. S2.2B shows the drift correction as a function of time over the 53-hour span of the measurement. Critically, we see that the drift undergoes a random walk, as we would expect from thermally dominated drift, and that over the course of 53 hours it only shifts by about 3 microns (approximately 10 pixels), an acceptable level of drift that can be corrected for without harm to the data.

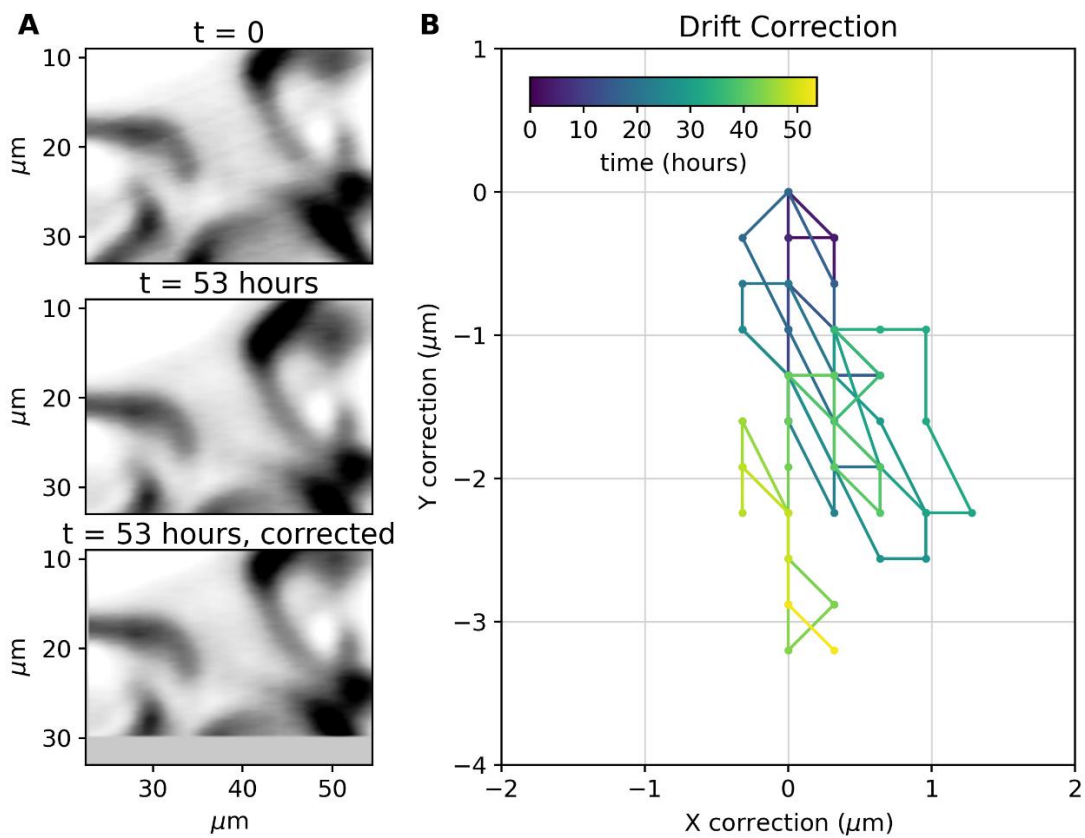

**Fig. S2.2** Drift correction for a 53-hour measurement imaging measurement. (**A**), shows the reflection image at the start of the measurement (top) and the end of the measurement (middle) and then the last image corrected to be spatially consistent with the first image (bottom). (**B**), the required drift correction as a function of time over the measurement.

## S3. Phonon Calculations, Band Structure and Phonon Amplitude Diagrams

### S3.1 Phonon calculation details

Vibrational coupling to the interlayer excitons at such large vibrational energies (~30 meV) suggests that the relevant vibrational modes are optical phonon modes. Equally spaced spectral sidebands (Fig. 2) indicate that the phonon *momentum* plays little role in the microscopic excitation process; only the conservation of energy is important. This is highly unusual in phonon-assisted processes since phonons usually provide both momentum and energy. In this case, the phonon momentum must be able to take any value in the microscopic formation of an interlayer exciton. This is evident in the fact that multiple sidebands are observed for *both* interlayer exciton species.

Combining these properties suggests that the phonon mode that couples to the exciton generation process must be an optical branch with energies near 30 meV that exhibits a flat band structure on the scale of the thermal energy $k_bT$. Flat phonon bands correspond to highly localized atomic vibrations, where each atom is an independent quantum harmonic oscillator, and each atom oscillates with the same frequency - the so-called Einstein mode. Using detailed phonon dispersion calculations (described below) we identify such phonon modes (Einstein modes) within the MoSe$_2$/WSe$_2$ heterostructure and explore the detailed properties of these modes. Remarkably, we find that the relevant modes exhibit flat bands with excitation energies near 30 meV.

The phonon dispersion of the heterostructure MoSe$_2$/WSe$_2$ with spin-orbit coupling (SOC) was calculated using the Vienna ab initio simulation package (VASP)[43-45] in the projected-augmented-wave method[46]. The generalized gradient approximation (GGA) with Perdew-Burke-Ernzerhof (PBE) parametrization[47-49] was used for the exchange correlation energy. The Van der Waals interactions between the layers are accounted by using the DFT-D2 Grimme's method[50]. The kinetic energy cutoff for our calculation is set as 500 eV.

For all structural relaxations, the convergence tolerance on the Hellmann-Feynman forces is less than $10^{-7}$ eV/Å. An 8×8×1 Γ-centered Monkhorst-Pack k-point mesh is used for 2D films. The vacuum layer added to the system is nearly 20 Å. Under these settings, the lattice constant of MoSe$_2$/WSe$_2$ is 3.3259 Å. The lattice mismatch between MoSe$_2$ and WSe$_2$ is less than 1%. Both of the two materials contain heavy elements; therefore, SOC is included for all calculations. The phonon dispersion is calculated for a 4×4×1 supercell using Phonopy[51].

## S3.2 Phonon dispersion relation and density of states

We first examine the phonon modes of this system, which Raman measurements indicate are strongly active near 30 meV (see main text Fig. 1A, B). As previously reported, there are a variety of Raman active and Raman inactive phonon modes near that energy[52]. To proceed we use the DFT calculations discussed in section S3.1 to model the dispersion of the phonon modes, and their density of states, shown in Fig. S3.1. Here, the faded blue bar marks 30 meV, but is broadened by $k_bT$ to represent the range of energies that would be consistent with the 30 meV mode observed in the data at $T = 20$ K. From Fig. S3.1, we see that there are multiple strongly active modes with bands near 30 meV. Each of the 27 bands in the dispersion corresponds to a particular mode of atomic vibration in the $WSe_2/MoSe_2$ heterostructure. Using the detailed results of our calculations, we next explore the atomic motions and band dispersions of each of these modes and identify which are most likely to be relevant to the absorption within the heterostructure. The right side of Fig. S3.1 shows the phonon density of states (DOS) versus phonon energy for the heterostructure, shown in black, which is sum of the DOS of the $WSe_2$ and $MoSe_2$ layers in blue and green respectively.

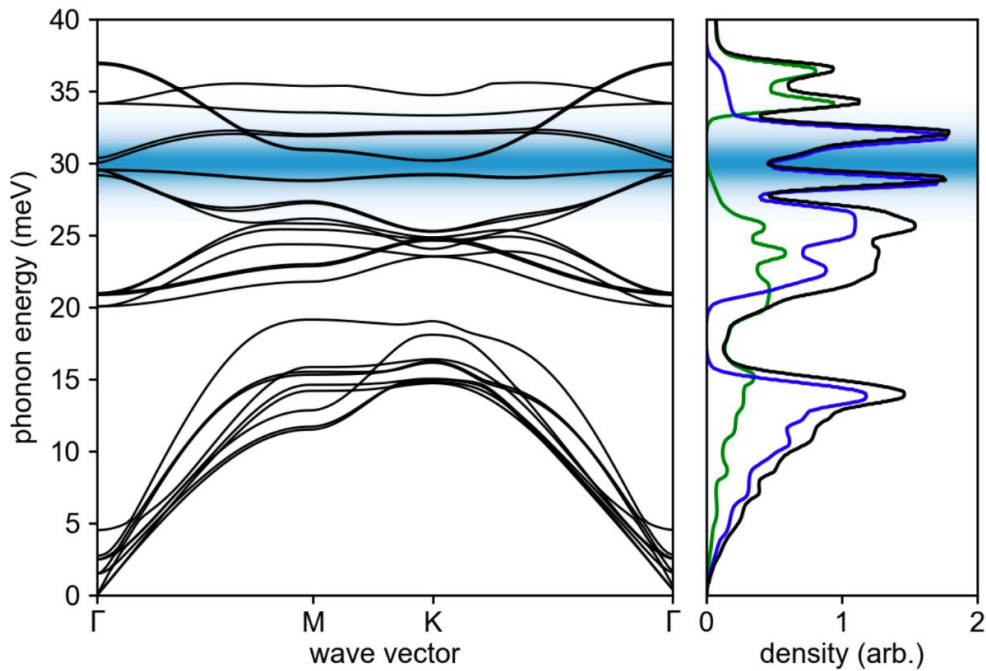

**Fig. S3.1** Dispersion of DFT calculated phonon modes (left), and, phonon density of states, for the $MoSe_2$ (green line), $WSe_2$ (blue line) and heterostructure (black line). Blue bar shows 30 meV with thermal broadening of $k_bT$ for $T = 20$ K (right).

Fig. S3.2 shows the displacements for the 7 modes at Γ near 30 meV corresponding to the 7 optical phonon branches shown in Fig. S3.3. The phonons mainly consist of in-plane vibrations, ($E_{2g}$ modes) with some out-of-plane vibrations ($A_{1g}$ modes). This indicates that there are several optical phonons, especially in-plane modes, near the energy of the oscillations we observe.

By comparing to our data, we identify candidate phonon modes which may be related to the photocurrent oscillations. In the low temperature data, Fig. 2, we see sharp peaks at 30 meV for photocurrent spectra involving transitions at different locations in the Brillion zone ($K \rightarrow K$ and $\Gamma \rightarrow K$). A vibrational (phonon) mode responsible for such oscillations must have an excitation energy of approximately 30 meV at both the $K$ and $\Gamma$ points.

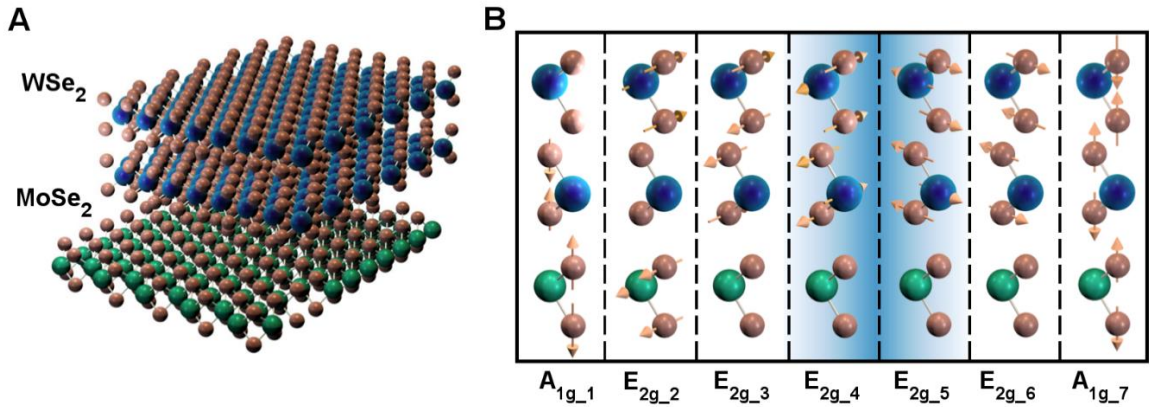

**Fig. S3.2** (**A**)**,** Device atomic structure geometry. (**B**), Visualization of optical phonons in the heterostructure along with their spectroscopy classification. Modes marked with the blue band correspond to nearly flat bands that are consistent with the oscillations observed in the photocurrent data.

In Fig. S3.3 we plot the seven phonon modes that are near 30 meV at the Γ point. We observe that most have a significantly different energy at the $K$ point, except for two which are marked as blue lines. These bands correspond to two nearly degenerate phonon modes, corresponding to the vibrations highlighted in Fig. S3.2B with the blue bars. Remarkably, in addition to exhibiting nearly dispersionless bands, these modes are polar optical modes, while the modes at neighboring energies are ordinary optical modes. We see that in these modes the tungsten and selenium atoms move oppositely in plane along either in-plane axis.

Having identified candidate phonon modes that may correspond to the 30 meV oscillation, phonon calculations confirm two critical inferences from the data. First, we note that these bands are almost flat, meaning they have nearly the same value at every wavevector. This would imply that they are not constrained by momentum; that a 30 meV phonon in these modes can contribute almost any momentum to the interlayer exciton excitation process. Second, we note that these polar phonon modes are localized on the same scale as the interlayer exciton. The interlayer exciton exhibits an electron localized in the $WSe_2$ layer, which vibrates in these modes. This implies that this phonon mode is vibrating the lattice over a comparable region to that of the exciton wavefunction, and that the interactions would be expected to be very strong owing to the polar nature of the phonons combined with the confined dipole moment of the interlayer exciton.

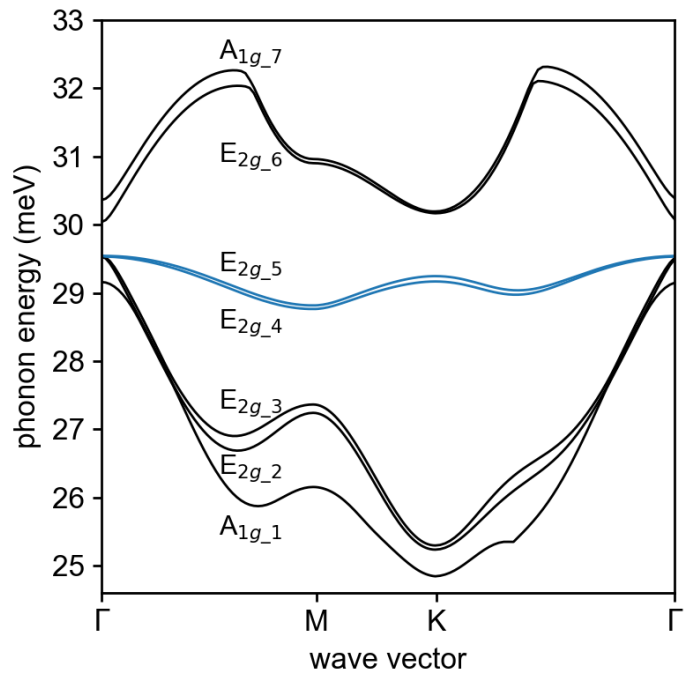

**Fig. S3.3** DFT calculated phonon dispersion for modes with Γ-point energies near 30 meV. Blue lines are the candidate optical phonon branches.

## S4. Additional Data Analysis: Photocurrent Spectroscopy

## S4.1. Photocurrent Spectroscopy of Intralayer Excitons

In this section we compare the photocurrent response of *intralayer* excitons to that of interlayer excitons within the vdW heterojunction. We show that the intralayer exciton photocurrent exhibits several key differences when compared to the interlayer photocurrent: First, photocurrent due to intralayer excitons *does not* exhibit regularly spaced peaks like those shown for interlayer excitons as in Fig. 2. Second, photocurrent due to intralayer excitons does not require the device to be tuned to precise charge neutrality and exhibits gate voltage dependence for positively and negatively charged trion species, consistent with ordinary behavior. We observe features that originate in both the $MoSe_2$ and $WSe_2$, with photocurrent excitation energies that match the relevant photoluminescence energies, consistent with ordinary photoresponse of intralayer excitons.

With high spectral resolution we tune the excitation energy close to the heterostructure peaks observed in the photoluminescence (Fig. 1B) and measure interlayer photocurrent. Fig. S4.1A shows photocurrent as a function of photon excitation energy from 1.55 eV to 1.37 eV (wavelength range between 800-900 nm), and $V_G$ between -20 to +20 V. We observed the appearance of three peaks at different photon energies that shift with gate voltage from $V_G$ = -20 V to +20 V. Notably, no sidebands are observed in the photoresponse of these peaks.

Fig. S4.1B shows the position of the peaks versus photon energy. We observed a peak at 1.53 eV, which we attribute to the neutral exciton ($X_{Mo}^0$) in $MoSe_2$. This peak does not shift as a function of gate voltage and corresponds approximately with the sharpest peak observed in the PL spectrum of $MoSe_2$, Fig. 1B. We attribute the photocurrent peak observed at 1.49 eV to the positively charged exciton (trion) of $MoSe_2$ (labeled $X_{Mo}^+$), which redshifts over the gate voltage range with a small slope, consistent with ordinary trion behavior. The photocurrent peak observed at 1.443 eV is attributed to the positive exciton (trion) $X_W^+$ originating from $WSe_2$. This peak again redshifts gradually as a function of gate voltage and corresponds closely to the PL peak position observed for $WSe_2$. We note that none of these peaks are observed exclusively at charge neutrality, instead exhibiting gate voltage dependence consistent with ordinary behavior.

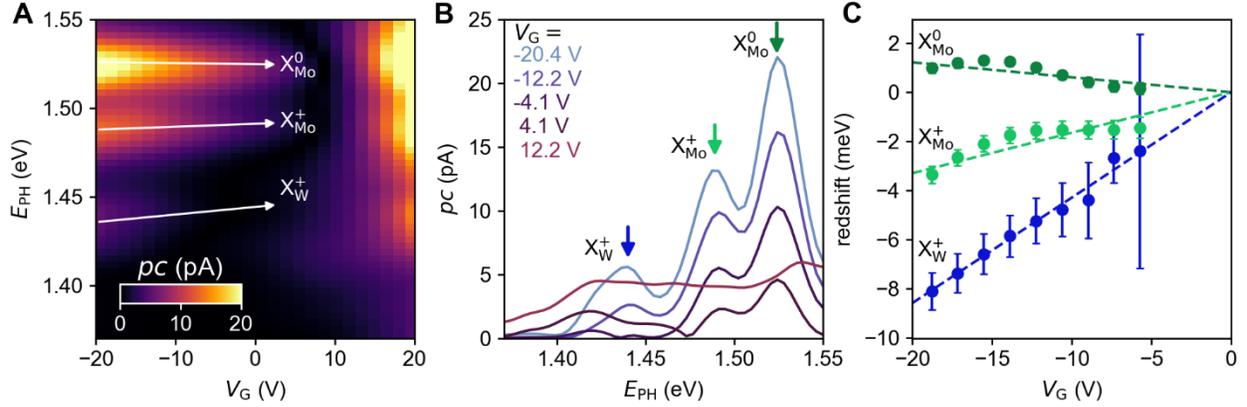

**Fig. S4.1** Gate voltage dependence of the excitation photon energy close to the energy of the intralayer excitons at room temperature. (**A**), Color map of photocurrent (dark current removed, see Section S.4.2.) versus excitation energy (1.37eV to 1.55eV) and $V_G$ at a fixed $V_{SD} = +6$ V ($T =$ 300 K). (**B**), Line cuts of interlayer photocurrent vs. excitation photon energy at several values of $V_G$ showing the existence of three peaks, the neutral exciton peak in MoSe$_2$ at 1.52eV, the positive exciton peak in MoSe$_2$ at 1.49eV, and the positive exciton peak in WSe$_2$ at 1.44eV. (**C**), Red shift of the peak positions versus gate voltage, neutral exciton in MoSe$_2$ (in dark green), positive exciton peak in MoSe$_2$ (in medium green), and positive exciton peak of WSe$_2$ (in blue). Peak positions calculated by fitting the photocurrent at constant $V_G$ to a function of three Lorentzian peaks, error bars indicate uncertainty in the fit. Dashed lines show a linear fit to the redshift extrapolated to $V_G = 0$ V.

**S4.2 Low Temperature Photocurrent Spectroscopy**

The low temperature photocurrent spectroscopy data shown in Fig. 2 was measured using the SSPSM setup described in section S2.1 on device 1 in an optical cryostat cooled with liquid helium to approximately 20 Kelvin. The spectral resolution of $E_{PH}$ is limited by a tradeoff in the SSPSM monochromator where increasing the spectral resolution decreases the pulse power (or equivalently, the signal to noise).

In the low temperature measurements, the dominant current signal is the dark current of the device with respect to $V_G$. Fig. S4.2A shows the raw current for the measurement shown in Fig. 2E (the Fig. 2A measurement is similar). We see that the raw signal is an order of magnitude larger than the photocurrent scale shown in Fig. 2E due to a large dark current. Fig. S4.2B shows that the photocurrent, i.e. the current varying with $E_{PH}$, is a small modulation on top of the dark current. There are several ways to process and remove the dark current, we use the simplest method, which is also advantageous for analyzing the photocurrent oscillations, that is we base our analysis on the derivatives with respect to $E_{PH}$, where the constant dark current drops out. For the purposes of displaying the data in Fig. 2A,E (and Fig. S4.3A,E) we set the minimum of the color scale at some threshold (shown as the grey dashed line in Fig. S4.2B) that displays the photocurrent in the most active region of $V_G$. This is purely visual, to give contrast to the photocurrent oscillations, and does not affect the subsequent analysis.

Fig. S4.3 breaks down the low temperature spectroscopy data in detail, showing line cuts of the data and the first and second derivatives of photocurrent with respect to $E_{PH}$ to visually display the photocurrent oscillations. The second derivative data in Fig. S4.3C, F was Fourier transformed to generate Fig. 2C, F.

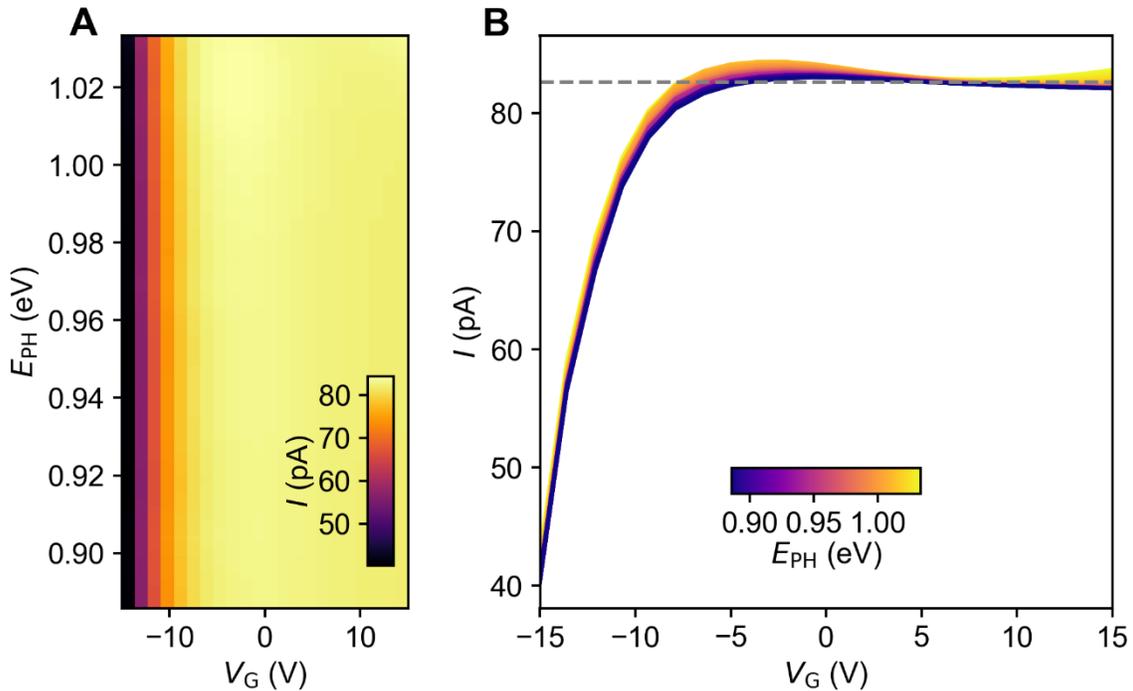

**Fig. S4.2** Raw low temperature photocurrent data. (**A**), raw Photocurrent, $I$ versus $E_{PH}$ and $V_G$, showing the primary effect from dark current. (**B**), raw photocurrent versus $V_G$ for variable $E_{PH}$ (shown as line color). Grey dashed line indicates the colorscale cutoff in Fig. 2E.

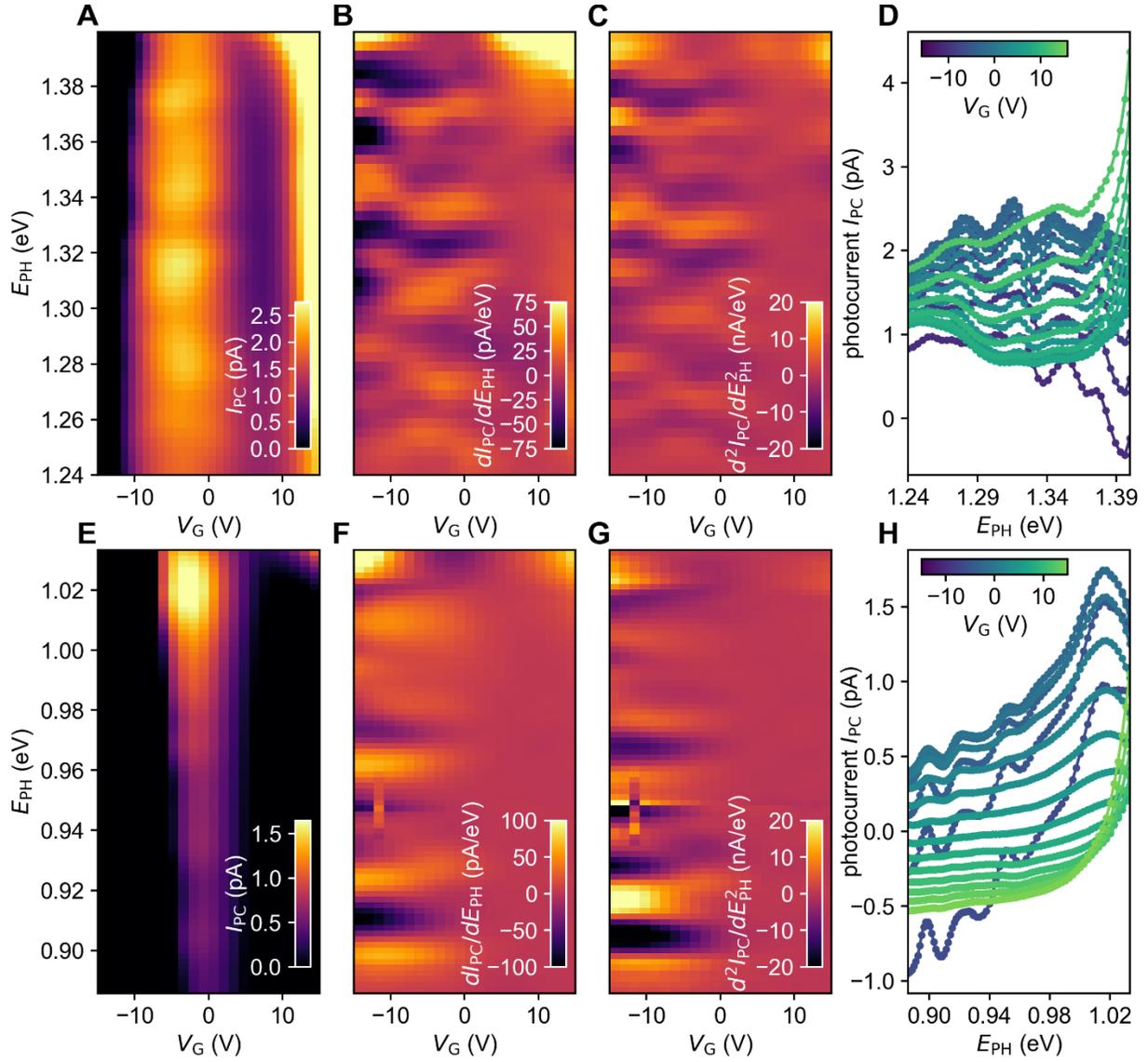

**Fig. S4.3** Low temperature photocurrent data. (**A**), Photocurrent, $I_{PC}$ versus $V_G$, near the 1.3 eV direct exciton along with its derivatives, (**B**), $dI_{PC}/dE_{PH}$ and, (**C**), $d^2I_{PC}/dE_{PH}^2$. (**D**) Line cuts of $I_{PC}$ versus $E_{PH}$ as a function of $V_G$. (**E**), Photocurrent, $I_{PC}$ versus $V_G$, near the 1 eV indirect exciton along with its first derivatives, (**F**), $dI_{PC}/dE_{PH}$ and it second derivative, (**G**), $d^2I_{PC}/dE_{PH}^2$. (**H**) Line cuts of $I_{PC}$ versus $E_{PH}$ as a function of $V_G$.

**S4.3 Data Intensive Exploration of Photocurrent Parameter Space**

In this section, we explore the spatial photocurrent response of device 2 in greater detail, showing the MPDPM imaging datasets used to generate the data show in Fig. 4. Fig. S4.4A shows a set of 2,000 photocurrent images versus $V_G$ and $V_{SD}$ with the laser wavelength at 1250 nm (1.0 eV). This is dataset is spatially consistent, i.e. corrected for spatial drift, and from the imaging set we can see the large-scale behavior of the device. Zooming in on a subset of the images in Fig. S4.4B and superimposing the outline of the heterostructure area of the sample, we see that the majority of the photocurrent originates outside the heterostructure, mainly from the semiconductor-metal interface at the device contacts, which will be discussed below. However, near $V_G = 0$ V we observe distinct and spatially uniform photocurrent originating from the heterostructure area, consistent with our expectation of photocurrent from an interlayer exciton near charge neutrality.

To verify that the photocurrent from the heterostructure is due to the interlayer exciton we take a similar set of spatial photocurrent maps at a fixed value of $V_{SD} = -0.35$ V and varying the laser wavelength, shown in Fig. S4.5A. Again, we identify the heterostructure region and consider the average photocurrent originating in that spatial area. The heterostructure responsivity, e.g. the photocurrent divided by the power, is shown in Fig. 4.5B left, and we see a bright stripe near $V_G = 0$ V. We average between $V_G = 0.0$ V and $V_G = 0.4$ V (right side of Fig. 4.5B) and observe a resonance in photon energy peaked at 0.96 eV, very close to our expectation of a 1.0 eV interlayer exciton.

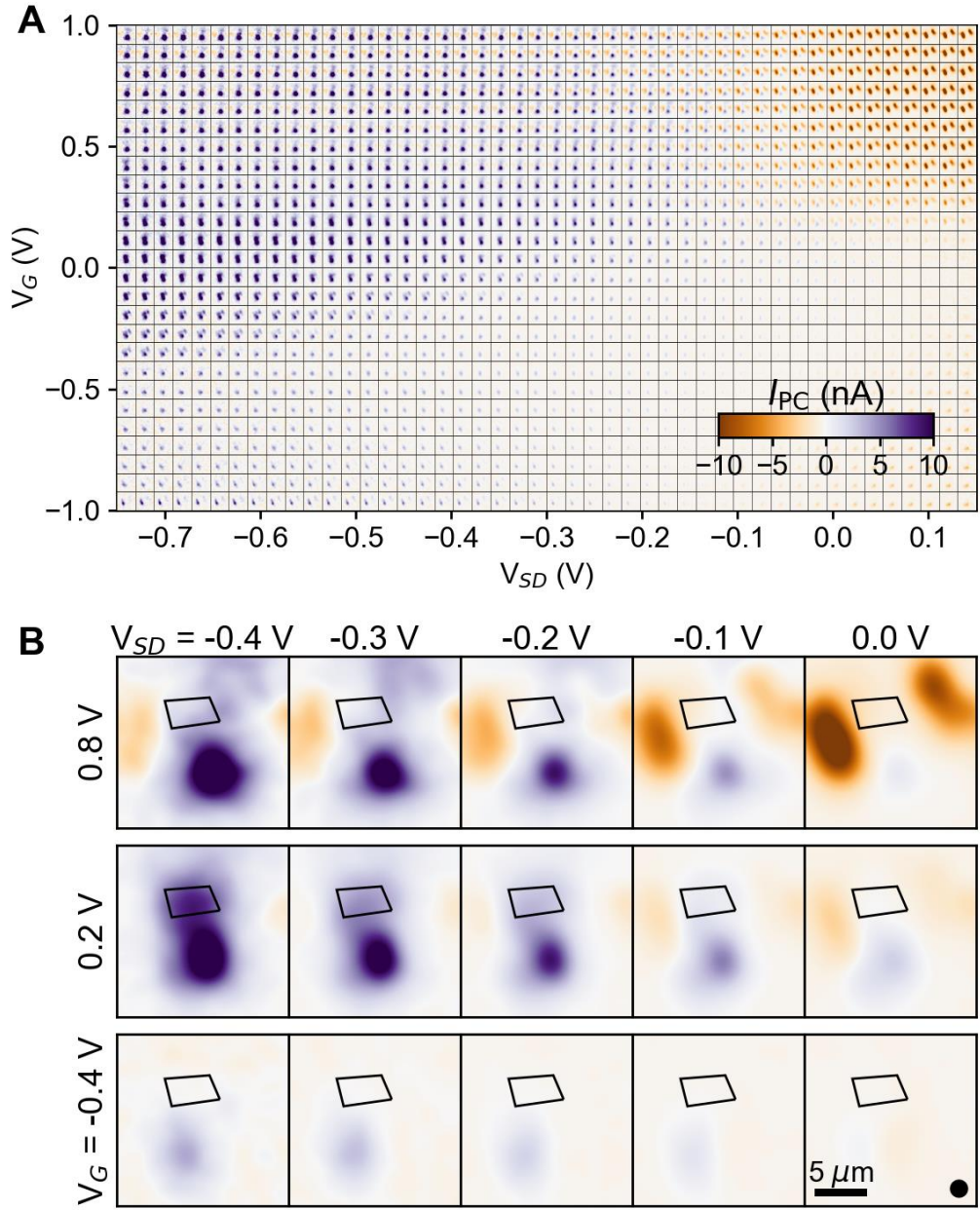

**Fig. S4.4** Data intensive imaging of the $V_G$ versus $V_{SD}$ parameter space. (**A**), data intensive imaging of device 2 sample at room temperature. (**B**), a series of photocurrent images showing the heterostructure area, outlined as a black rectangle.

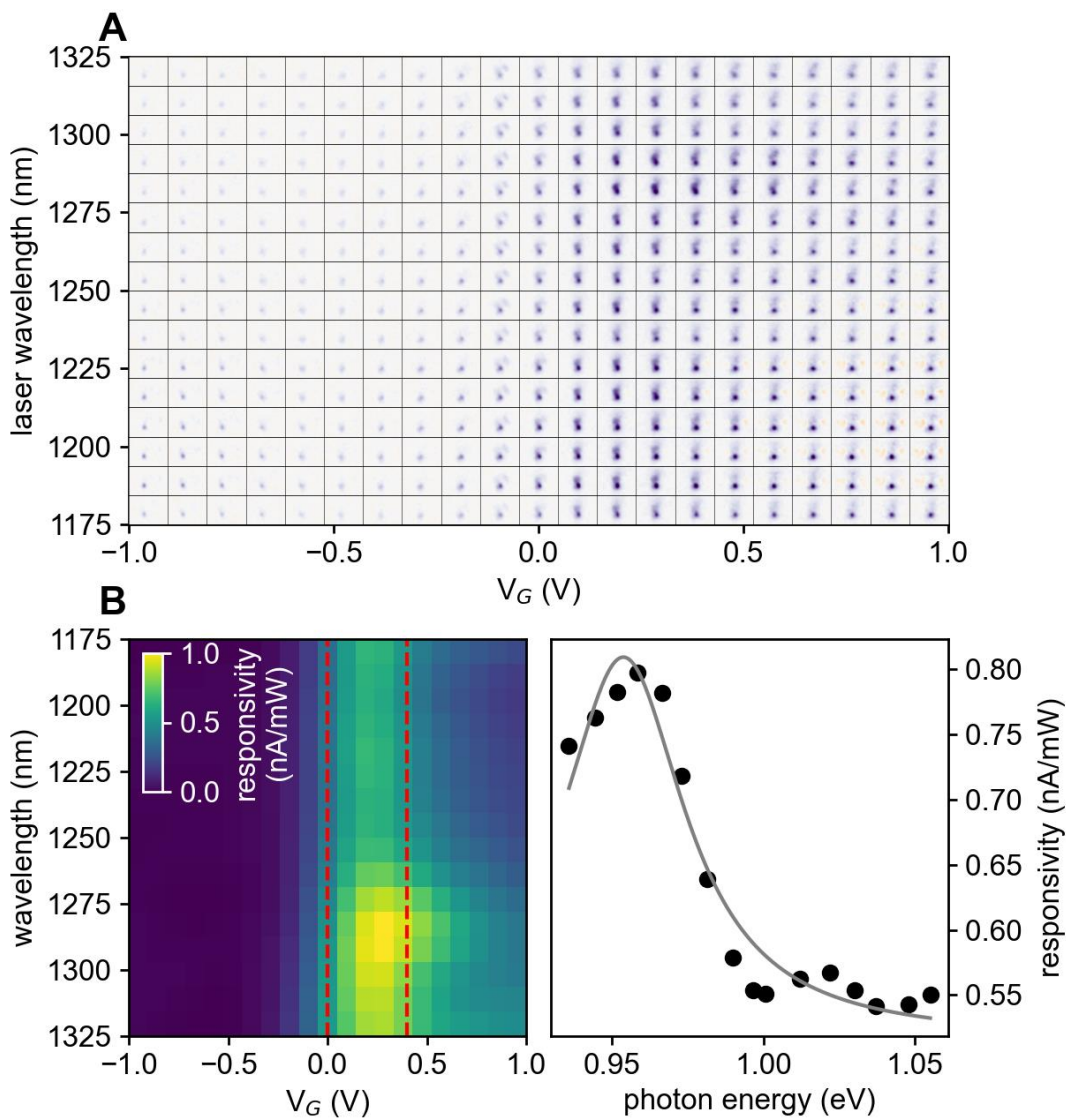

**Fig. S4.5** Data intensive imaging of the $V_G$ versus laser wavelength parameter space. (**A**), data intensive imaging of device 2 showing photocurrent versus wavelength and gate voltage. (**B left**), processed responsivity (photocurrent divided by laser power) from the heterostructure regions and, (**B right**), the averaged responsivity (points) fit to a Lorentzian function (grey line).

There are multiple spatially distributed sources of photocurrent in the imaging data sets, the majority of which come from the metal contacts. Fig. S4.6 visualizes the photocurrent at the three points in space corresponding to the principle sources of photocurrent, $WSe_2$ contact, the heterostructure and the $MoSe_2$ contact. Fig. S4.5A compares a photocurrent map and a reflection image and identifies the locations of the photocurrent as points. The heterostructure photocurrent in Fig. S4.6C has been well described in this work. The $WSe_2$ and $MoSe_2$ photocurrent in Fig.

S4.6B and S4.6D are both more active for positive values of $V_G$ and occur opposite to each other with respect to $V_{SD}$. Photocurrent at the contacts results from the photo-thermoelectric effect at the metal-semiconductor interface, where the laser produces a local temperature gradient between two materials with significantly different Seebeck coefficients[53].

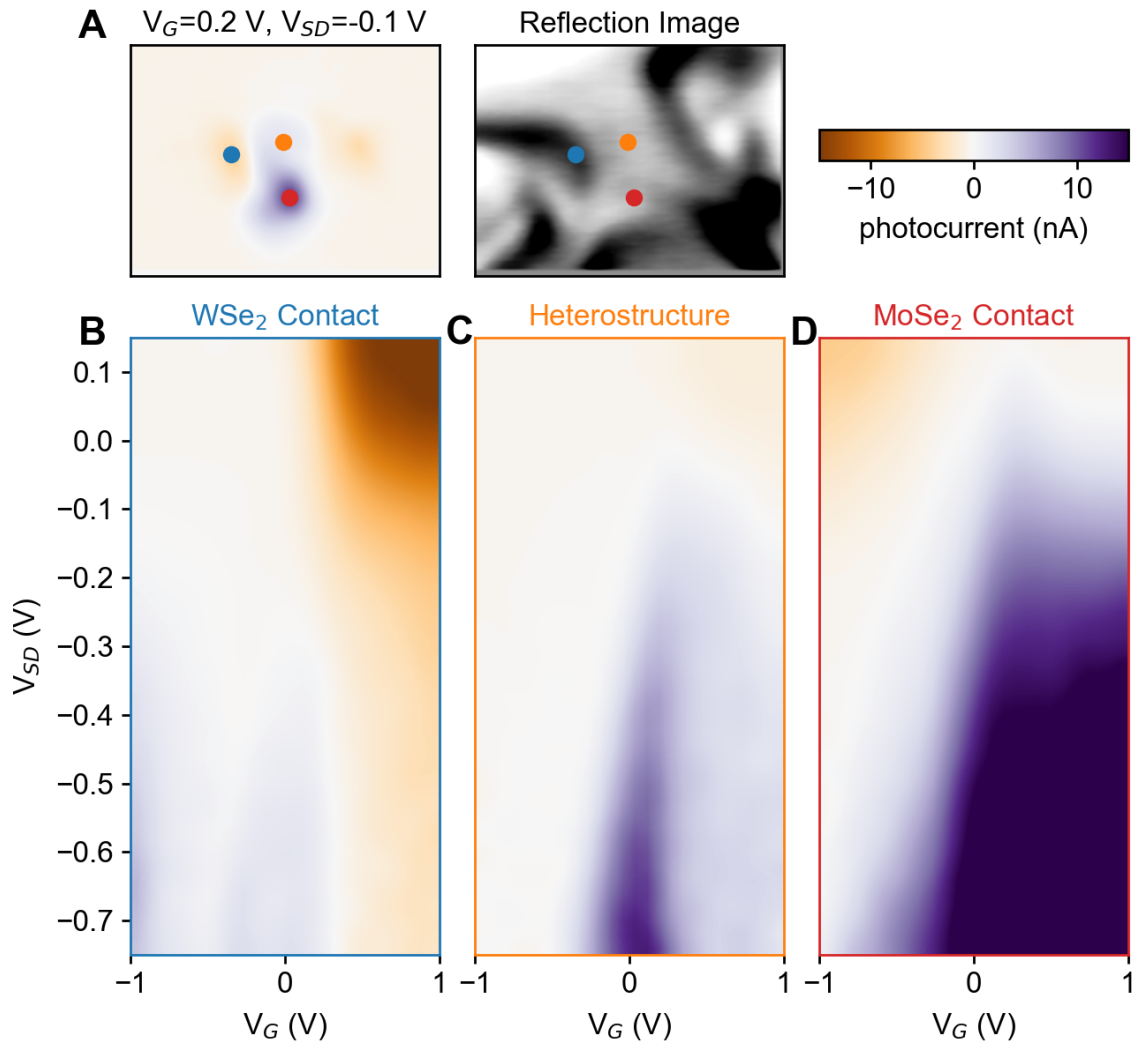

**Fig. S4.6** Photocurrent from the metal contacts versus the heterostructure. (**A**), Example colormap (colorscale exaggerated) compared to the corresponding reflection image. (**B-D**) the photocurrent from the WSe$_2$ contact, the heterostructure contact and the MoSe$_2$ contact respectively, marked in a with the blue, orange and red points.

## S4.4. Analysis of Photocurrent Oscillations in the Current-Voltage Characteristics

In the main text, we show that the energy change due to an effective interlayer voltage $V_I$ can be written as $\Delta E = \vec{p} \cdot \vec{E} = ed\, V_I / \eta t_{TMD}$ where $|\vec{p}| = ed$ is the dipole moment of the interlayer exciton with length $d$, $t_{TMD}$ is the heterostructure thickness, and $\eta > 1$ reduces the effective voltage due to depletion capacitance. Furthermore, we relate changes in the interlayer electric field energy to a reduced voltage $eV_{SD}/\alpha$, giving $\Delta E \approx e\,V_I/\eta = e\,V_{SD}/\alpha\eta$. By carefully analyzing the p-n junction behavior of the device, we are able to re-scale the current-voltage characteristics to give insight into the electric field energy experienced by the interlayer exciton within the heterostructure. In this section, we provide additional detailed analysis of the photocurrent vs. $V_{SD}$ and $V_G$, and extract the empirical parameters $\alpha$ and $\eta$.

In Fig. S4.7A we plot the average heterostructure photocurrent versus $V_{SD}$ and $V_G$ and observe a bright feature of positive photocurrent near $V_G = 0$ as a function of increasing $V_{SD}$. To proceed, we will look at the photoconductance, that is $dI_{PC}/dV_{SD}$, which we plot in Fig. S4.7B. We observe a bright feature corresponding to the main feature, but we also see a number of small oscillations in the conductance. To further visualize these oscillations, we take the second derivative $d^2I_{PC}/dV_{SD}^2$, shown in Fig. S4.7C, where we see strong oscillations in the photocurrent.

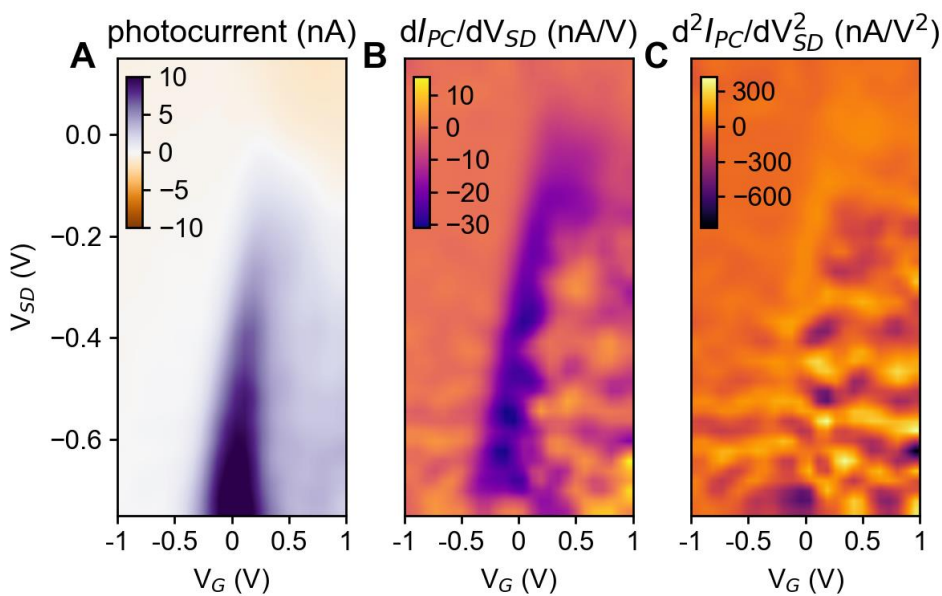

**Fig. S4.7** Analysis of the $V_{SD}$ versus $V_G$ parameter space. (**A**), heterostructure photocurrent $I_{PC}$ of device 2 versus $V_{SD}$ and $V_G$, calculated by averaging the images in Fig. S4.4 within the heterostructure area. The first derivative $dI_{PC}/dV_{SD}$ (**B**) and second derivative $d^2I_{PC}/dV_{SD}^2$ (**C**) with respect to $V_{SD}$.

Furthermore, in main text Fig. 4B the energy shift across the heterostructure is obtained by scaling the voltage $V_{SD}$ by two factors: $\alpha$ and $\eta$. $\alpha$ is obtained from fitting the dark current in Fig. 4A to the diode equation, $I = I_s(e^{eV_{SD}/\alpha k_b T} - 1)$ (shown in Fig. 4A), while $\eta$ can be determined by the interdependence of the photocurrent on $V_{SD}$ and $V_G$. By analyzing the heterostructure photocurrent, we find that discretely spaced peaks in the photocurrent spectra emerge only if the device is finely tuned to charge neutrality. The interlayer photocurrent extracted from $2 \times 10^3$ photocurrent images (shown in Fig. S4.4A) occurs predominantly near $V_G = 0$ V, increases as $V_{SD}$ increases in reverse bias ($V_{SD} < 0$ V), and follows the diagonal trajectory (solid line in Fig. S4.8A).

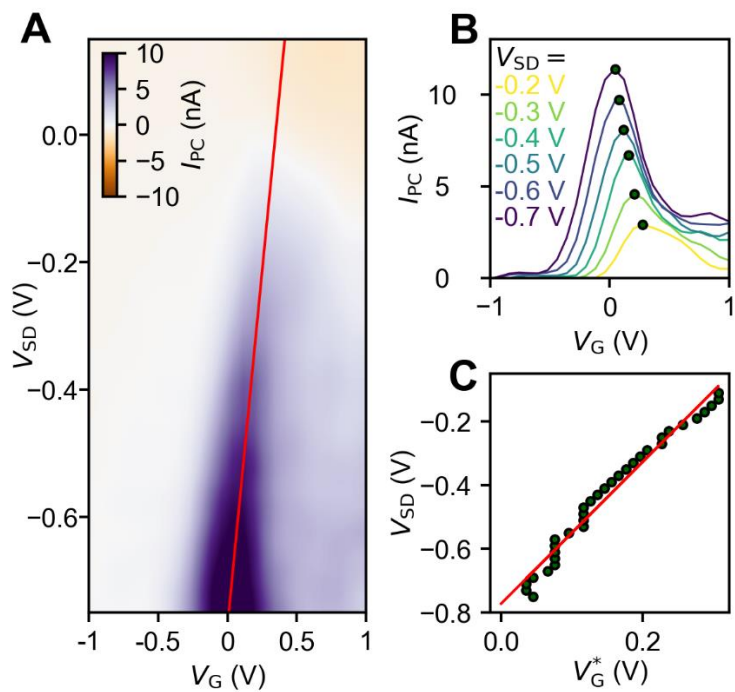

**Fig. S4.8** Identifying diagonal self-capacitance behavior in the photocurrent parameter space. (**A**), the heterostructure photocurrent parameter space which mainly consists of a bright feature that is diagonal with respect to $V_{SD}$. (**B**), line cuts of the photocurrent versus $V_{SD}$, green points identify the maxima of the curves. (**C**), shows the $V_G$ value of the photocurrent maximum $V_G^*$ versus $V_{SD}$. The red line is the linear relation between $V_G^*$ and $V_{SD}$, which we see tracks the diagonal of the photocurrent in (**A**).

By plotting the $V_G$-dependent characteristics taken at several $V_{SD}$ values (Fig. S4.8B), we extract the gate voltage values of the photocurrent peaks $V_G^*$, and plot $V_G^*$ as a function of $V_{SD}$ (Fig. S4.8C). In addition to increasing the electric field across the WSe$_2$-MoSe$_2$ interface, the effective interlayer voltage $\Delta V_I \propto \Delta V_{SD}$ contributes to charging the individual layers, giving rise to a depletion capacitance[54]. At charge neutrality, the charge density $n = 0$ and the change in gate voltage $\Delta V_G$ can be directly related to the change in $\Delta V_I$: $\Delta V_I = -(C_G/C_I)\Delta V_G$, where $C_I$ and $C_G$ are the capacitances per unit area for the interlayer depletion region and due to the bottom gate, respectively[55]. From a linear fit to the data we find that changes in $V_{SD}$ are indeed proportional to changes in $V_G$: $\Delta V_{SD} = -\eta \Delta V_G \approx -2.1 \Delta V_G$. Along the diagonal line in Fig. S4.7A, the interlayer electric field increases while maintaining precise charge neutrality. The value of $\eta$ is obtained from fitting $V_G^*$ as a function of $V_{SD}$. This linear relationship, shown as red lines in Fig. S4.8A and Fig. S4.8C, arises due to the self-capacitance and the slope of the line is $\eta$. The effective interlayer voltage is reduced as energy is committed to charging the interface, giving rise to an interlayer capacitance. While this effect is well-known in semiconductor heterojunctions, our analysis is the first to identify it atomically thin p-n junctions.

Having now converted the period of room temperature voltage oscillations into units of energy across the heterostructure, we can compare it to the low temperature photocurrent spectroscopy data. Fig. S4.9 compares the low temperature photocurrent FFTs (from Fig. 2C, G) and the room temperature imaging FFT (from Fig. 4C) with the phonon density of states (from Fig. S3.1). We observe that in all cases there is a strong oscillation corresponding to the 30 meV phonon. Remarkably this oscillation is still observable, and thus exciton localization persists, even at room temperature.

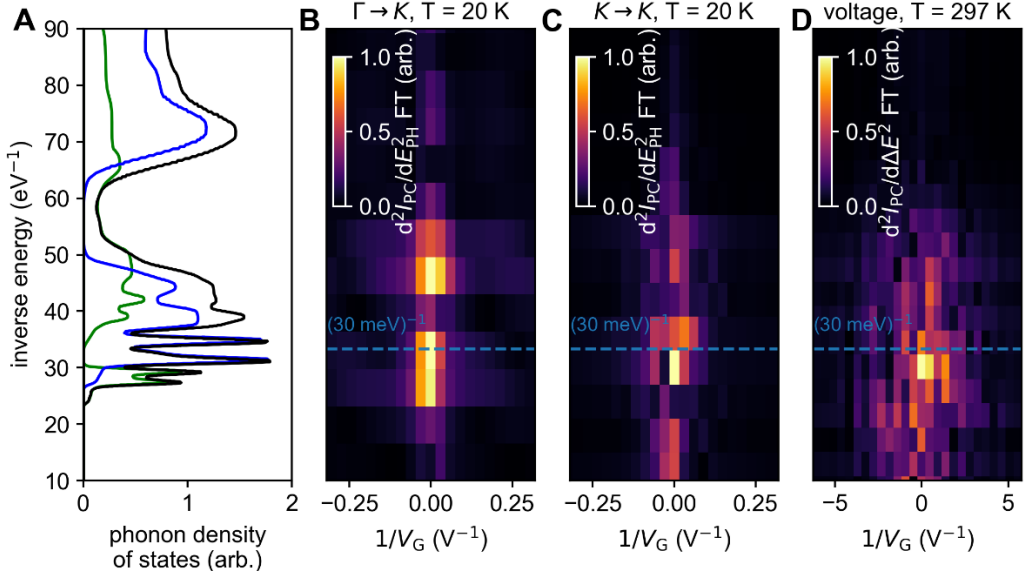

**Fig. S4.9** Comparison of photocurrent oscillations across multiple regimes. (**A**), the phonon density of states from Fig. S3.1 plotted as a function of inverse energy for comparison with Fourier transforms. (**B**) and (**C**), Fourier transforms of the second derivative of the heterostructure photocurrent of device 1, labeled by the exciton transition they measure. (**D**), Fourier transform of the second derivative of the heterostructure photocurrent of device 2. In all cases there is a bright mode at (30 meV)$^{-1}$ marked with a blue dashed line.

The photocurrent oscillations observed in Fig. 4 and discussed above do not occur in the dark current of the system, confirming that they originate from the interlayer exciton. The photocurrent signal is measured using a lock-in amplifier to reject the large dark current. To confirm that the lock-in is functioning properly, and that the oscillations originate from the photocurrent we compare the photocurrent and photocurrent spectrum to the dark current. Fig. S4.10A shows the second derivative of the photocurrent data from Fig. 4 and its Fourier transform and Fig. S4.10B shows the second derivative of the dark current over a similar range of Δ$E$ and its Fourier transform. Looking at the Fig. S4.10B we note that the dark current is dominated by high frequency noise and is not correlated with the 30 meV oscillation clearly observed in the photocurrent. Thus, we conclude that the oscillation originates from the photocurrent, and therefore the exciton, and not the dark current.

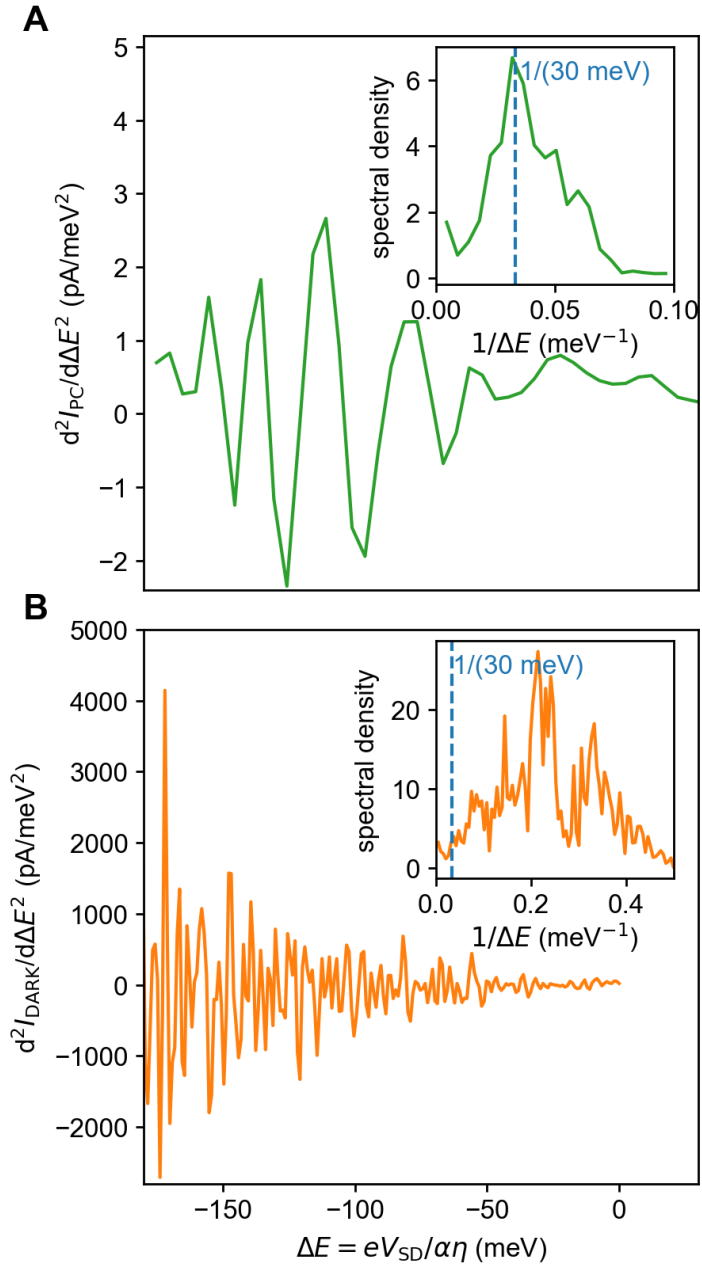

**Fig. S4.10** Comparison of photocurrent oscillations to the dark current. (**A**) Second derivative of heterostructure photocurrent and its Fourier transform (**inset**) from Fig. 4B. (**B**) Second derivative of dark current, measured with no laser light over a similar voltage range and it's Fourier transform (**inset**).

**S5 Franck-Condon Transitions and Strong Exciton-Phonon Coupling**

In this section, we discuss Franck-Condon transitions, the multiple phonon sidebands, as well as the conditions under which strong exciton-phonon coupling is manifested.

**S5.1 Franck-Condon Transitions and Sidebands**

We first review how multiple sidebands arise from Franck-Condon transitions and strong exciton-phonon coupling. We begin by analyzing an exciton transition from a ground state, $g$, to an excited state, $e$, where the excited state is strongly coupled with the atomic configuration of the system (captured by a phonon coordinate, $Q$). For simplicity of presentation, we consider a single (optical) phonon mode with frequency $\omega_0$, and its coupling to a localized exciton mode where exciton-phonon coupling is most pronounced, see below for a discussion of localized vs delocalized excitons. The energy surfaces for the ground and excited states can be described by Ref. 56

$$E_g(Q) = \frac{1}{2} A Q^2, \qquad E_e(Q) = \varepsilon_0 + \frac{1}{2} A (Q - Q_0)^2, \qquad (S1)$$

where $\varepsilon_0$ is the bare energy of the transition from ground to excited state in the absence of exciton-phonon coupling, $A = M\omega_0^2$ is an elastic constant and $M$ is an effective mass characterizing the phonon mode. Here $Q_0$ encodes the relative atomic displacement between the energy minima of the electronic ground and excited state manifolds, which results from the exciton-phonon coupling. For instance, when $Q_0 = 0$, there is no exciton-phonon coupling (Fig. 3A of the main text), yielding an absorption that is characterized by a single broad absorption peak. In contrast, when $Q_0 \neq 0$, exciton-phonon coupling can lead to multiple side-band features in the absorption (Fig. 3B of the main text).

The multiple side-band features can be understood heuristically (see also explicit derivation below) by noting that the atomic configuration, characterized by the phonon coordinates, does not change in the short time over which an electronic transition takes place (the so-called "Franck-Condon principle")[57,58]. This means that optical transitions are "vertical" in the phonon coordinates. Since the energy surfaces are displaced by $Q_0$, optical transitions originating from the minimum of the electronic ground state energy surface to the electronic excited state energy surface can leave the phonon mode in a range of excited states, thereby producing multiple side-bands. The strength of this effect is captured by the characteristic number of phonons involved in the transition, i.e., the average excitation number in the phonon state after the vertical transition. This quantity is characterized by the dimensionless Huang-Rhys factor, $S = A Q_0^2/(2\hbar\omega_0)$. Interestingly, $S$ also characterizes the depth of the potential well, $\Delta = S\hbar\omega_0$, resulting from the displaced phonon coordinates in the excited state (displaced oscillator).

We now explicitly account for these multiple side-band transitions by considering the vibronic states involved in the optical excitation. The ground and excited state manifolds are characterized by the vibronic states[56]:

$$\text{Ground: } |\psi_g^X, \phi_n\rangle, \quad \varepsilon_g(n) = (n + 1/2)\hbar\omega_0, \quad (S2)$$
$$\text{Excited: } |\psi_e^X, \tilde{\phi}_m\rangle, \quad \varepsilon_e(m) = \varepsilon_0 + (m + 1/2)\hbar\omega_0, \quad (S3)$$

where $|\phi_n\rangle$ are the modes of the *undisplaced* oscillator (optical phonon mode) in the electronic ground state, while $|\tilde{\phi}_m\rangle$ corresponds to the modes of the *displaced* oscillator in the electronic excited state. Here $n, m = 0, 1, 2, 3, ...$ are integers, and $|\psi_{g,e}^X\rangle$ are the ground and excited excitonic wavefunctions respectively (For simplicity we assume that the excitonic states $|\psi_{g,e}^X\rangle$ are independent of $n, m$, as typical for the Franck-Condon principle)[58].

Crucially, because the oscillator modes in the excited state $|\tilde{\phi}_m\rangle$ are displaced (in coordinate space, by $Q_0$) from the those in the undisplaced oscillator of the ground state, their overlap can be non-vanishing even when $m \neq n$. Indeed, the overlap between $|\tilde{\phi}_m\rangle$ and $|\phi_0\rangle$ produces the familiar Poisson distribution:

$$|\langle\tilde{\phi}_m|\phi_0\rangle|^2 = (S^m/m!) \exp(-S). \quad (S4)$$

As a result, the rate of optical transitions in the presence of incident light with energy $\hbar\omega$ follows directly from the golden rule:

$$\Gamma_{g\to e}(m, \hbar\omega) = \frac{2\pi}{\hbar}|\langle\psi_e^X, \tilde{\phi}_m|H'|\psi_g^X, \phi_0\rangle|^2\delta(\varepsilon_e(m) - \varepsilon_g(0) - \hbar\omega)$$
$$= \Gamma_0 (S^m/m!) \exp(-S) \, \delta(E_0 + m\hbar\omega_0 - \hbar\omega), \quad (S5)$$

where $\Gamma_0 = (2\pi/\hbar)|\langle\psi_e^X|H'|\psi_g^X\rangle|^2$ is the bare transition rate in the absence of exciton-phonon coupling and $H'$ is the light-matter interaction. This allows for phonons to be involved in the electronic transition between $g$ and $e$, producing the multiple sidebands as shown in the main text. We can estimate the value of $S$ from the range over which the multiple sidebands persist in the photocurrent spectra. For example, given that numerous photocurrent sidebands manifest (about 7) in Fig. 2, we estimate $S \sim 2 - 3$. This yields a potential well depth (caused by the exciton-phonon coupling) of up to $\sim 90$ meV.

**S5.2 Localized Excitons vs Delocalized Excitons**

Strong exciton-phonon coupling - and its concomitant pronounced sidebands - is a hallmark of localized exciton systems. To appreciate this, it useful to recall that the lowest excitonic mode in an extended crystal is the zero center of mass kinetic energy mode. This is a delocalized exciton formed as a superposition in which the bound electron-hole pair is coherently spread out throughout the system. The kinetic energy of the exciton center of mass degree of freedom has a characteristic energy bandwidth $2w$. As a result, the creation of a localized exciton requires mixing many plane waves, i.e. states of different center of mass momentum, and thus comes with a kinetic energy cost of order $w$.

Interaction with the lattice provides a mechanism to lower the energy of a localized exciton and can compete with the kinetic energy cost described above. It does this by relaxing the lattice around it (rearranging the atoms around the exciton), lowering its energy by $\Delta$ (see above section). When $\Delta > w$, the exciton can become self-trapped and is localized[7,9]. In such a localized state, the absorption spectrum of the exciton can exhibit multiple side-bands as described above (e.g., exhibiting the multiple peaks seen in the photocurrent spectra of the main text). The appearance of multiple and pronounced sidebands are hallmarks of strong exciton-phonon coupling and a localized exciton mode.

In contrast, when $w > \Delta$, the exciton will prefer to be in the delocalized state, freely propagating through the crystal[5,7,9]. Its absorption spectrum is dominated by the broad absorption peak typical of delocalized excitons (see, e.g., illustration in Fig. 3A of main text).